\documentclass[9pt, twocolumn]{article}
\usepackage[left=0.75in,right=0.75in,top=1in,bottom=1in]{geometry}
\usepackage{amsfonts}
\usepackage{footnote}
\usepackage{hyperref}
\usepackage{titling}
\usepackage{tikz}
\usepackage{amsmath}
\usepackage{algorithm}
\usepackage{algorithmic}
\usepackage{appendix}
\usepackage{graphicx}
\usepackage{caption}
\usepackage{subcaption}
\usetikzlibrary{patterns}
\usepackage{pgfplots}
\usepackage{tablefootnote}
\usepackage{enumitem}
\usepackage{array}
\usepackage{amsthm}
\usepackage{amsmath}
\usepackage{xspace}
\usepackage{multirow}
\usepackage{lipsum}
\usepackage{titlesec}
\titlespacing*{\section}{0pt}{10pt}{5pt} 
\titlespacing*{\subsection}{0pt}{10pt}{5pt}
\titlespacing*{\subsubsection}{0pt}{10pt}{5pt} 

\newtheorem{proposition}{Proposition}[section]
\newtheorem{definition}{Definition}[section]

\newcommand{\ourmodel}{DPOT}
\newcommand{\myparatight}[1]{\smallskip\noindent{\bf {#1}:}~}
\newcolumntype{M}[1]{>{\centering\arraybackslash}m{#1}}
\newcolumntype{x}[1]{>{\centering\arraybackslash\hspace{0pt}}p{#1}}
\newcommand{\eqnref}[1]{(\ref{#1})\xspace}

\AtBeginDocument{%
  }

    \title{\vspace{-1cm}Concealing Backdoor Model Updates in Federated Learning by Trigger-Optimized Data Poisoning}
    
    \author{Yujie Zhang \\ Duke University \\ yujie.zhang396@duke.edu \and Neil Gong  \\ Duke University \\ neil.gong@duke.edu \and Michael K. Reiter \\ Duke University \\ michael.reiter@duke.edu}
    \date{\vspace{-4mm}}

\begin{document}
    
\maketitle
\begin{abstract}
Federated Learning (FL) is a decentralized machine learning method
that enables participants to collaboratively train a model without
sharing their private data. Despite its privacy and scalability
benefits, FL is susceptible to backdoor attacks, where adversaries
poison the local training data of a subset of clients using a backdoor
trigger, aiming to make the aggregated model produce malicious results
when the same backdoor condition is met by an inference-time input.
Existing backdoor attacks in FL suffer from common deficiencies: fixed
trigger patterns and reliance on the assistance of model
poisoning. State-of-the-art defenses based on analyzing clients' model updates exhibit a good defense performance on these attacks
because of the significant divergence between malicious and benign client model updates. To effectively conceal malicious model updates among
benign ones, we propose \ourmodel{}, a backdoor attack strategy in FL
that dynamically constructs backdoor objectives by optimizing a
backdoor trigger, making backdoor data have minimal effect on model
updates. We provide theoretical justifications for \ourmodel{}'s
attacking principle and display experimental results showing that
\ourmodel{}, via only a \textit{data}-poisoning attack, effectively
undermines state-of-the-art defenses and outperforms existing backdoor
attack techniques on various datasets.
\end{abstract}

\section{Introduction}

Federated Learning (FL) is a decentralized machine-learning approach
that has gained widespread attention for its ability to address
various challenges. Unlike traditional centralized model training, FL
enables model updates to be computed locally on distributed devices,
offering enhanced data privacy, reduced communication overhead, and
scalability for a large number of clients.  In each round of FL, a central server
distributes a global model to participating clients, each of whom
independently trains the model on their local data, and their model
updates are aggregated by the server for updating the global model.

Despite its advantages, FL has been proven susceptible to backdoor
attacks~\cite{howtobackdoor}. Backdoor attacks in federated learning involve adversaries inducing the local models of a subset of clients to learn backdoor information carried by triggers and strategically integrating these backdoored local models into the global model. Ultimately, the global model will generate the adversary-desired result when the
same trigger conditions are met. In this work, we term clients manipulated by adversaries during local training as \textit{malicious clients}, and those unaffected as \textit{benign clients}.

Existing backdoor attacks in FL present two common
deficiencies. First, the patterns of backdoor triggers are pre-defined
by the attacker and remain unchanged throughout the entire attack
process~\cite{howtobackdoor, Xie2020DBA:, sun2019can,
  gong2022coordinated}. Consequently, the optimization objective
brought by backdoored data (backdoor objective) is static and
incoherent with the optimization objective of main-task data (benign
objective), resulting in distinct differences in model updates after
training. These malicious clients' model updates are therefore easily
canceled out by robust aggregations~\cite{howtobackdoor,
  zhang2022neurotoxin,gong2022coordinated}. Second, many approaches
rely on model-poisoning techniques to enhance the effectiveness of backdoor
attacks. 
Implementing
model-poisoning attacks requires attackers to change the training
procedures of a certain number of genuine clients (e.g., at least 20\%
of all clients~\cite{alittleisenough, fang2020local,yes}) to make
their local training algorithms different from other clients.
However, achieving this condition is challenging, as advanced defense
mechanisms~\cite{rieger2022crowdguard} have introduced Trusted
Execution Environments (TEEs) to ensure the secure execution of
client-side training, making it harder to adopt suspicious
modifications to the training procedure.

Existing defenses against backdoor attacks in FL (see more details in
Section~\ref{sec:defense}) rely on a hypothesis that backdoor attacks
will always cause the updating direction of a model to deviate from
its original benign objective, because the backdoor objectives defined
by backdoored data cannot be achieved within the original
direction. However, the capabilities of backdoor attacks are not
limited to this hypothesis.  To counter this hypothesis, adversaries
can align the updating directions of a model with respect to backdoor
and benign objectives by strategically adjusting the backdoor
objective. Applying this idea to FL, if the injection of backdoored
data has minimal effect to the updates of a client's model, then
detecting this client as malicious becomes challenging for defenses
based on analyzing clients' model updates.

In this work, we propose \textbf{D}ata \textbf{P}oisoning with \textbf{O}ptimized \textbf{T}rigger (\ourmodel{}), a backdoor attack on FL that
dynamically constructs the backdoor objective to to continuously minimize the divergence between clients' model updates in the backdoored states and the non-attacked states. We construct the backdoor objective by
optimizing the backdoor trigger that is used to poison malicious
clients' local data. Without any assistance of model-poisoning
techniques, malicious clients can effectively conceal their model
updates among benign clients' model updates by simply executing a
normal training process on their poisoned local data, and render
state-of-the-art defenses ineffective in mitigating our backdoor
attack.

The optimization of the backdoor trigger in each round is independent
and specific to the current round's global model. The objective of
this optimization is to generate a trigger such that the current
round's global model exhibits minimal loss on backdoored data having
this trigger. Once the global model becomes optimal for the backdoored
data, further training on the backdoored data will result in only
minor model updates to the current state of global model within a limited number of
local training epochs.  Therefore, when a malicious client's local
dataset is partially poisoned by the optimized trigger while the rest
remains benign, the model updates produced by training on the local
data will be dominated by benign model updates. We provide both theoretical and experimental
justifications for the sufficiency of trigger optimization in
minimizing the difference between a malicious client's model updates
in the backdoored state and the non-attacked state.

In order to enhance the practicality of our attack,
we limited the trigger size to a reasonable level, ensuring it cannot
obscure essential details of the original data. To meet this constraint, we developed two
algorithms to separately optimize trigger pixels' placements and values. To the best of our knowledge, we are
the first to generate an optimized trigger with free shape and
placement while specifying its exact size.

We evaluated \ourmodel{} on four image data sets (FashionMNIST,
FEMNIST, CIFAR10, and Tiny ImageNet) and four model architectures
including ResNet and VGGNet. We assessed the attack effectiveness of
\ourmodel{} under a variety of defense conditions, testing it against 10
defense strategies that are based on analyzing clients' model updates --- these include Median~\cite{trim_median}, Trimmed Mean~\cite{trim_median}, RobustLR~\cite{robustlr}, Robust Federated Aggregation (RFA)~\cite{RFA}, FLAIR~\cite{flair}, FLCert~\cite{cao2022flcert}, FLAME~\cite{nguyen2022flame}, FoolsGold~\cite{foolsgold}, Multi-Krum~\cite{krum}, and FRL~\cite{everyvotecounts}  --- along with one defense strategy that uses client-side adversarial training to recover the global model: Flip~\cite{zhang2022flip}. We compared \ourmodel{} attack with three state-of-the-art data-poisoning backdoor attacks that
employ fixed-pattern triggers, distributed fixed-pattern triggers (DBA~\cite{Xie2020DBA:}),
and partially optimized triggers (A3FL~\cite{zhang2024a3fl}), respectively.  Using a small number of
malicious clients (5\% of the total), \ourmodel{} outperformed
existing data-poisoning backdoor attacks in effectively undermining
defenses without affecting the main-task performance of the FL system.

In summary, our contributions are as follows:
\vspace{-\topsep}
\begin{itemize}[leftmargin=*]
  \setlength{\itemsep}{0pt}
  \setlength{\parskip}{0pt}
  \setlength{\parsep}{0pt}
\item We propose a novel backdoor attack mechanism, \ourmodel{}, in FL
  that effectively conceals malicious client's model updates among those of
  benign clients by dynamically adjusting backdoor objectives, and
  demonstrate that existing defenses focusing on analyzing clients'
  model updates are inadequate.
\item We dynamically construct the backdoor objective solely by optimizing the
  backdoor trigger and injecting it to clients' data (a.k.a. data-poisoning), without relying on additional assistance from model-poisoning techniques.
\item We offer both theoretical and experimental justifications for the
  adequacy of our trigger optimization in reducing the disparity
  between model updates in the backdoored state and the non-attacked state.
\item We develop algorithms to optimize a trigger, allowing for
  flexibility in its shape, placement, and values, while precisely constraining
  its size.
\item We extensively evaluate our attack on four benchmark datasets,
  showing that \ourmodel{} achieves better attack effectiveness than
  three advanced data-poisoning backdoor attacks in compromising 11
  state-of-the-art defenses in FL.
\end{itemize}

\vspace{-2mm}
\section{Related Work}
\begin{figure*}[htbp] 
  \centering
  \small
\[
  \text{\parbox{3.5cm}{\centering \textbf{ \large Backdoor Attacks \\ in FL}}} 
  \left\{
  \begin{array}{l}
    \text{\parbox{1.5cm}{\centering With \\ model \\ poisoning}} 
    \left\{
   \begin{array}{l}
      \text{\parbox{2.4cm}{\centering Static objective \\ (fixed trigger)}} 
      \left\{
        \begin{array}{l}
          \text{Semantic trigger: ~\cite{howtobackdoor}} \\
          \text{Edge-case backdoor:~\cite{yes}} \\
          \text{\parbox[c][1cm][c]{2.5cm}{Artificial trigger}} 
          \left\{
            \begin{array}{l}
              \text{\parbox[c][0.3cm][c]{5cm}{Single global trigger:~\cite{sun2019can,zhang2022neurotoxin,alittleisenough} }} \\
              \text{\parbox[c][0.3cm][c]{5cm}{Distributed trigger: ~\cite{Xie2020DBA:}}}
            \end{array}
          \right. 
        \end{array}
      \right. \\ 
      \text{\parbox{2.8cm}{\centering Dynamic objective \\ (optimized trigger)}} 
      \left\{
        \begin{array}{l}
          \text{$L_2$-norm bounded trigger: ~\cite{CerP}} \\
          \text{\parbox[c][1cm][c]{2.3cm}{$L_0$-norm \\ bounded trigger}} 
          \left\{
            \begin{array}{l}
              \text{\parbox[c][0.5cm][c]{6cm}{Fixed shape and placement: ~\cite{fang2023vulnerability}}} 
            \end{array}
          \right. 
        \end{array}
      \right.
    \end{array}
    \right.\\
    \\
    \text{\parbox{1.5cm}{\centering Without \\ model \\ poisoning}} 
    \left\{
      \text{\parbox{3cm}{\centering Dynamic objective \\ (optimized trigger)}} 
      \left\{
        \begin{array}{l}
          \text{\parbox[r][1cm][c]{2.3cm}{$L_0$-norm \\ bounded trigger}} 
          \left\{
            \begin{array}{l}
              \text{\parbox[c][0.3cm][c]{6cm}{Fixed shape and placement: ~\cite{gong2022coordinated,zhang2024a3fl}}} \\
              \text{\parbox[c][0.3cm][c]{6cm}{Free shape and placement: \textbf{Our work}}}
            \end{array}
          \right. 
        \end{array}
      \right.
    \right.
    \end{array}
  \right.
  \]
  \vspace{-4mm}
  \caption{An overview of related works on backdoor attacks in FL.}
  \label{tree:attacks}
  \vspace{-3mm}
\end{figure*}

\subsection{Federated Learning (FL)} 

The Federated Learning~\cite{mcmahan2017fl} (FL) training process involves four main steps: 1)~\textbf{Model
  Distribution}: A central server distributes the most recent global
model to the participating clients. 2)~\textbf{Local Training}: Each client independently trains the global model on its local training dataset and obtains a local model.
 3)~\textbf{Model Updates}: Each client calculates the parameter-wise difference between its local model and the global model, referred to as model updates, and then sends them to the central server.   4)~\textbf{Aggregation}: The central server aggregates clients' model updates to create a new global model.  This entire process, consisting of step
1 to 4, constitutes a global round.  The FL system repeats these steps for a certain number of rounds to obtain a final version of the global model.

\subsection{Backdoor Attack}

Backdoor attack in machine learning is a security vulnerability where
an adversary manipulates a model's behavior by making it
learn some trigger information, causing the model to produce 
erroneous results when trigger conditions are met.  Meanwhile, the backdoor attack also ensures that the model maintains normal performance on benign data to evade abnormal detection.  In image classification tasks, a backdoor attack aims to manipulate a model so that it classifies any image containing a specific pixel-pattern trigger into a label chosen by the attacker
\cite{chen2017targeted, gu2019badnets, kolouri2020universal,
  lin2020composite, yao2019latent, wang2019neural, liu2018trojaning}.

\subsection{Backdoor Attacks in FL}
\vspace{-1mm}

FL is easily suffered from backdoor attacks. As training data are privately held by clients, the security of data is hard to track or protect. Adversaries can inject backdoors into the global model simply by compromising a few vulnerable client devices and poisoning their data with backdoor triggers. To date, many variations of backdoor attacks targeting FL have emerged, and we summarize those specific to image classification tasks in Figure~\ref{tree:attacks}.  

\myparatight{With model poisoning v.s. Without model poisoning} 

The foundation of backdoor attacks in FL is through \textbf{\textit{data poisoning}} - attackers embed backdoor triggers into the local training data of certain clients and change the ground-truth labels of the infected data to malicious labels. As a result, clients' local models trained on the poisoned data will be backdoored, and consequently, the global model that aggregates these backdoored models will also be backdoored.

A standalone data poisoning is found challenging to succeed when employing some types of triggers. Therefore, many works introduce model poisoning to assist backdoor attacks in FL. \textbf{\textit{Model poisoning}} aims to either directly manipulate clients' model updates or indirectly achieve this by changing their local training algorithms. Three main approaches in model poisoning were widely adopted in existing attacks: 1) Scaling based~\cite{howtobackdoor,sun2019can,Xie2020DBA:,gong2022coordinated}. Attackers amplify malicious model updates generated from backdoored models before clients send them to the server. These malicious updates can overpower the aggregation results, causing the global model to quickly incorporate backdoors. However, this approach is vulnerable to defenses that exclude outlier model updates from the aggregation. 2) Constraint based~\cite{howtobackdoor,CerP}. Attackers change clients' local training algorithms by adding extra constraints to their loss functions, giving backdoored models specific characteristics, such as being less distinguishable from benign models.  3) Projection based~\cite{zhang2022neurotoxin,alittleisenough,yes,fang2023vulnerability}. Attackers constrain backdoor implementation to bounded model parameters: by clipping parameter values or using Projected Gradient Descent, backdoor models are $L_2$-norm bounded to a chosen model state; by selectively updating a subset of parameters, they are $L_0$-norm bounded to a chosen state.

Model poisoning requires attackers to modify certain clients' local training procedures. However, with the introduction of Trusted Execution Environments (TEEs) by state-of-the-art defense mechanisms~\cite{rieger2022crowdguard}, client-side execution for training can be authenticated and secure, thus increasing the difficulty of conducting model poisoning.  In contrast, data poisoning is easier to conduct and harder to prevent since clients may collect their local data from open resources where attackers can also get access to and make modifications. For example, autonomous driving
vehicles collect their data on road traffic signage~\cite{xie2022efficient} from real world, and attackers can easily place stickers on traffic signage objects to inject backdoor trigger information. Hence, we consider backdoor attacks that do not involve model poisoning to be more advanced than those that do.

\myparatight{Static objective v.s. Dynamic objective}

\begin{figure*}[htbp] 
  \begin{center}
  \small
\[
  \text{\parbox[c][3cm][c]{3.5cm}{\centering \textbf{\large Defenses against \\ Backdoor Attacks \\ in FL $^\dag$}}} 
    \left\{
   \begin{array}{l}
        \text{\parbox{2.8cm}{ (By server) \\ Analyzing clients' \\ model updates}}
          \left\{
            \begin{array}{l}
              \text{\parbox[c][1cm][c]{8cm}{Excluding model updates with outlier values: \\ \textit{\small FLAIR~\cite{flair}, FoolsGold~\cite{fung2018mitigating}, RobustLR~\cite{robustlr}, FLAME~\cite{nguyen2022flame}}} }\\
              \text{\parbox[c][1cm][c]{9cm}{Excluding model updates that are outliers in certain features: \\ \textit{\small BayBfed~\cite{baybfed}, FRL~\cite{everyvotecounts}, FreqFed~\cite{freqfed}}}} \\
              \text{\parbox[c][1.3cm][c]{8cm}{ Byzantine-robust aggregation: \\ \textit{\small Trimmed Mean~\cite{trim_median}, Median~\cite{trim_median}, Multi-Krum~\cite{krum}, RFA~\cite{RFA}, FLCert~\cite{cao2022flcert} }}}
            \end{array}
          \right.  \\
      \text{\parbox{10cm}{ (By clients) \\Recovering by adversarial training: \textit{\small Flip~\cite{zhang2022flip}}}}
    \end{array}
    \right.
  \]
\end{center}
\vspace{-3mm}
  \rule{0in}{1.2em}$^\dag$ \footnotesize In this work, we only discuss defenses that adhere to the fundamental privacy-preserving principles of FL~\cite{mcmahan2017fl} - clients' private data are kept local, and their model updates are not shared with any entities other than the server.
  \vspace{-2mm}
  \caption{An overview of related works on defenses against backdoor attacks in FL.}
  \label{tree:defenses}
  \vspace{-3mm}
\end{figure*}

If a backdoor attack has a specified and unchanging objective that is independent to the training system's status, we refer to this as a \textbf{\textit{static objective}}. For instance, Semantic trigger as backdoor~\cite{howtobackdoor} aims to associate certain features from input that is unrelated to the main training tasks with an attacker-chosen output, causing the model to make incorrect predictions on those inputs; Edge-case backdoor~\cite{yes} selects data that share certain commonalities but are from the tail end of the input data distribution as the backdoored input, causing the model to mispredict them; Artificial trigger as backdoor~\cite{sun2019can,zhang2022neurotoxin,alittleisenough,Xie2020DBA:} embeds a few pixels forming a specific artificial pattern into the input, leading the model to mispredict any input containing this pixel pattern. In FL, since the static objectives of backdoor attacks are inconsistent with the optimization objectives defined by the main-task data, malicious models will exhibit distinct differences in their model updates compared to benign models, making them easy to detect.

In contrast to a static objective, a backdoor attack that adjusts its objective based on the training system's status is referred to as having a \textbf{\textit{dynamic objective}}. By adjusting its objective, a backdoor attack is expected to achieve greater effectiveness. Several approaches have been proposed in recent attack studies to attempt to accomplish this. For example, Model-dependent attack~\cite{gong2022coordinated} and F3BA~\cite{fang2023vulnerability} optimized the trigger pattern based on a hypothesis that maximizing the activation of certain neurons in the backdoored local model can enhance the attack's persistence on the global model, which is however lack of theoretical evidence and proof-of-concept codes; A3FL~\cite{zhang2024a3fl} optimized triggers specifically for a corner case in FL training, where the global model is directly trained to unlearn the trigger, but the effectiveness of A3FL triggers in more general FL training scenarios remains unaddressed.

\myparatight{$L_2$-norm bounded optimized trigger v.s. $L_0$-norm bounded optimized trigger}

A critical consideration in designing backdoor triggers is ensuring their stealthiness when applied to input data, resulting in a substantial disparity between human perception and the backdoored model's interpretation. Existing dynamic objective attacks achieve this by constraining the optimized triggers' $L_2$-norm or $L_0$-norm bounds.

An \textbf{\textit{$L_2$-norm bound}} on a trigger or perturbation means that the total magnitude of the changes introduced by the backdoor is limited. This makes the perturbation subtle, ensuring it doesn't drastically alter the input data. For example, CerP~\cite{CerP} generates optimized perturbations of the same size as a data point for each round and adds them to clients' local data to induce their local models learn to misclassify the perturbed data to a specified target label. 

An \textbf{\textit{$L_0$-norm bound}} restricts the number of components (e.g., pixels in an image) that can be altered by the trigger. This constraint ensures that the trigger is sparse, meaning it only affects a small portion of an input data. For example, optimized triggers in Model-dependent attack~\cite{gong2022coordinated}, F3BA~\cite{fang2023vulnerability}, and A3FL~\cite{zhang2024a3fl} all consist of a small number of pixels arranged in a square shape and are placed in a fixed corner location on the data to poison. 

An $L_2$-norm bounded trigger is less practical for real-world data poisoning because it spreads changes across many pixels, requiring the attacker to access and alter a figure's values before it is physically printed for use. Additionally, these small perturbations are easily disrupted by data preprocessing techniques that filter out unnecessary noise. In contrast, an $L_0$-norm bounded trigger is easier to apply to any data (e.g., a sticker on an image) due to its stable shape, consistent values, and compact size. However, existing works in optimizing $L_0$-norm bounded triggers are limited by fixing their shapes and placements and only updating triggers' values, which fails to fully leverage the potential of optimized triggers for attacking FL.

\subsection{Defenses against Backdoor Attacks in FL}\label{sec:defense}

In this work, we focus on discussing defenses that adhere to the fundamental privacy-preserving principles of FL introduced by McMahan, et al~\cite{mcmahan2017fl} - clients' private data are kept local, and their model updates are not shared with any entities other than the server. We summarize the related defense works in figure~\ref{tree:defenses}. For a discussion on additional defenses with varying privacy-preserving properties, please refer to the Appendices~\ref{apx:related_work}.

In existing defenses, the server and clients are the two subjects commonly considered for implementing defense strategies. For clients as the defense subject, the global model of each round is the input they receive from the FL system. Flip~\cite{zhang2022flip} proposed using trigger inversion on the global model and adversarial training on local models to mitigate the impact of the backdoor trigger, which is a defense strategy implemented by benign clients. However, Flip's effectiveness against optimized triggers remains unaddressed. Optimized triggers are more challenging to recover than fixed ones due to their variability across different rounds. For the server as the defense subject, clients' model updates are the input that the server receives from the FL system.
Numerous studies have proposed defenses against backdoor attacks by analyzing clients' model updates, which can be further classified into three categories, as discussed below.

\myparatight{Excluding model updates with outlier values}
Some existing works believed that a malicious client's model updates will directly exhibit significant difference in values from those of benign clients, therefore excluding model updates with outlier values can mitigate the effects of backdoor attacks. FLAME~\cite{nguyen2022flame} and FoolsGold~\cite{foolsgold} exclude a client's model updates that have outlier cosine similarity in values to other clients' model updates. FLAIR~\cite{flair} and RobustLR~\cite{robustlr} reduce or penalize the contribution of model updates that show a certain degree of sign dissimilarity, either on a client-wise or element-wise basis.

\myparatight{Excluding model updates that are outliers in certain features}
Some existing works believed that the effects of backdoor attacks could be reflected on some features extracted from model updates' values, so they proposed to filter model updates according to certain features. BayBfed~\cite{baybfed} and FreqFed~\cite{freqfed} assess the probabilistic distribution and frequency transformation of clients' model updates, and eliminate outlier clients based on these features. FRL~\cite{everyvotecounts} creates a sparse space of model updates for clients to vote, where the server rejects outlier votings and aggregates the acceptable updates within this space.

\myparatight{Byzantine-robust aggregation}
Some existing works propose aggregating only the most trustworthy model updates to tolerate the presence of malicious clients, which we refer to as byzantine-robust aggregation. Median and Trimmed Mean~\cite{trim_median} aggregate reliable model updates element-wise, while Multi-Krum~\cite{krum} and RFA~\cite{RFA} select and aggregate reliable model updates client-wise. FLCert~\cite{cao2022flcert} takes the majority inference results from the reliable models of different client groups to mitigate the influence of malicious clients.

Analyzing clients' model updates can effectively defend against backdoor attacks with static objectives due to the great divergence existing between malicious clients' and benign clients' model updates. However, when a backdoor attack can dynamically change its objective to effectively eliminate the difference between a client's model updates in malicious state and benign state, defenses based on this strategy may struggle to succeed.

\section{Threat Model}\label{threat}
\input{draw}

\myparatight{Attacker's capability} As shown in
Figure~\ref{fig:principle}, we assume that each FL client---even a
malicious one---is equipped with trustworthy training software that
conducts correct model training on the client's local training data
and transmits the model updates to the FL server. Aligning with the
security settings in the state-of-the-art defense
work~\cite{rieger2022crowdguard}, we assume that both the client
training pipeline and the FL server, as well as the communication
between them, faithfully serve FL's main task training and cannot be
undetectably manipulated.  These properties would be achievable by
executing FL training within Trusted Execution Environments
(TEEs)~\cite{schneider:2022:sok,rieger2022crowdguard}, for example, by
applying cryptographic protections to the updates (e.g., a digital
signatures) to enable the FL server to authenticate the updates as
coming from the TEEs.

Due to the TEE's protections, malicious clients are not allowed to
conduct any model poisoning.  The capability of
malicious clients in our attack is limited to the manipulation of
their local training data that are input to their training
pipelines---i.e., a data-poisoning attack.  In addition, in line with existing works~\cite{CerP,zhang2024a3fl,fang2023vulnerability,gong2022coordinated}, we do not assume the secrecy of the global model provided by the FL server, as it would typically need to be accessible outside TEEs for use in local inference tasks.  As such, in each FL round, clients are granted white-box access to the global model.

\myparatight{Attacker's background knowledge} In our attack, we
consider the presence of malicious clients in the FL system. As
discussed above, malicious clients can white-box access to the global
model in each round. Originating from initially benign clients that have been compromised, these malicious clients possess some local training data for the FL main task.

\myparatight{Attacker's goals} The malicious clients aim to accomplish
the following goals.
\begin{itemize}[leftmargin=*]
  \vspace{-2mm}
  \setlength{\itemsep}{0pt}
  \setlength{\parskip}{0pt}
  \setlength{\parsep}{0pt}
\item {\bf Effectiveness}.  By convention, \textit{Attack Success
  Rate} (ASR) is used to assess the effectiveness of a backdoor
  attack. For classification tasks, ASR is defined as the accuracy of
  a model in classifying data embedded with a backdoor trigger into the target label associated with this trigger.
  The \ourmodel{} attack aims to cause the global model in each FL round to misclassify data embedded with a backdoor trigger, generated in the previous round, into a target label. Our effectiveness goal is for the global model to achieve an ASR of over 50\% in the final round and even maintain an average ASR of over 50\% across all rounds.

\item {\bf Stealthiness}.  The stealthiness goal of a backdoor attack
  is to maintain the \textit{Main-task Accuracy} (MA) of the global
  model at a normal level, ensuring the functionality of the global
  model on its main-task data.  Specifically, we require that the
  compromised global model resulting from our attack has a similar MA
  ($\pm$ 2 percentage points) compared to a global
  model that has not been subjected to any attacks.
\end{itemize}

\section{\ourmodel{} Design}
\subsection{Overview}

In each round of FL (e.g., the $i$-th round), \ourmodel{} attack takes
place after the malicious clients receive the global model ${W_g}^{(i)}$
of this round but before they input their local training data to the
trustworthy training software. Given a global model ${W_g}^{(i)}$ and a
pre-defined target label $y_t$, we optimize the pattern of a backdoor
trigger to increase the $\mathit{ASR}$ of ${W_g}^{(i)}$. By poisoning malicious
clients' local training data using this optimized trigger
$\tau^{(i)}$, we expect that the global model of the next round
${W_g}^{(i+1)}$ will also exhibit a high ASR in classifying data
embedded with $\tau^{(i)}$ into its target label $y_t$.

To achieve this goal, we first construct a trigger training dataset by
collecting data from malicious clients. After changing the labels of
all data in the trigger training dataset to be the target label $y_t$,
we compute the gradient on each pixel of each image with respect to
the loss of
${W_g}^{(i)}$ in misclassifying each clean image into lable $y_t$. We
determine trigger-pixel placements $E_t$ within the pixel location space of an image by
selecting $\mathit{tri}_{\mathit{size}}$ number of pixels that demonstrate largest absolute
values among the pixel-wise sum of gradients from all images. Next, we
optimize each pixel's value in $E_t$ using gradient descent, obtaining
the trigger-pixel values $V_t$. Finally, we embed the optimized trigger
defined by ($E_t$, $V_t$) with its target label $y_t$ into malicious
clients' local training data at a certain poison rate.

\subsection{Building a Trigger Training Dataset}

At the beginning of the \ourmodel{} attack, we initially gather all
available benign data from the malicious clients' local training
datasets and assign a pre-defined target label $y_t$ to them. We refer
to this new dataset, which associates benign data with the target
label, as the trigger training dataset $D$.

\subsection{Optimizing Backdoor Trigger}

\myparatight{Formulating an optimization problem} We use the trigger
training dataset to generate a different backdoor trigger for each
round's global model. The optimization process operates independently
across the rounds of FL, implying that generating a backdoor trigger
for the current round's global model does not depend on any
information from previous rounds. Therefore, in this part, we
introduce the trigger optimization algorithms within a single round of
FL.

In the image classification context, consider the global model $W_g$
as input and all pixels within an image as the parameter space. Our
approach aims to find a subset of parameters that have the most
significant impact in producing the malicious output result (i.e.,
target label), and subsequently optimize the values of the parameters
in this subset for the malicious objective (i.e., a high ASR). In the
end, the pixels in this subset with their optimized values will serve
as a backdoor trigger. This trigger will increase the likelihood that
an image containing it will yield the malicious output when employing
the same model $W_g$ for inference. The optimization objective to
resolve the above problem can be written as formula \ref{eq:algobj}.
\begin{align}\label{eq:algobj}
    \min_{\tau }\quad& \frac{1}{\mid D \mid}\sum_{x \in D}\mathit{Loss}(W_g(x \odot \tau), y_t), 
\end{align}
where $\tau$ represents the backdoor trigger composed of trigger-pixel placements
 $E_t$ and trigger-pixel values $V_t$. The objective is to minimize
the difference between the target label $y_t$ and the output results of
the global model $W_g$ when taking the backdoored images as input,
which can be quantified by a loss function. The symbol $\odot$
represents an operator to embed the backdoor trigger $\tau$ into a
clean image $x$, whose definition is further described in
\eqnref{eq:operator} of Section~\ref{sec:thry}. To enhance
generalization performance of the optimized backdoor trigger, we
employ all images in the trigger training dataset $D$ as constraints
and try to find a backdoor trigger that takes effect for all of these
images.

\myparatight{Solving the optimization problem} To solve the above
optimization problem, our approach employs two separate algorithms:
one for computing the trigger-pixel placements $E_t$ (see
Algorithm~\ref*{alg:triloc}), and the other for optimizing the trigger-pixel values $V_t$ using $E_t$ as input (see Algorithm~\ref*{alg:trival}).

{\bf Compute trigger-pixel placements $E_t$.} In
Algorithm~\ref*{alg:triloc}, we select pixel locations that contain
the largest absolute gradient values with respect to the backdoor
objective \eqnref{eq:algobj} as the trigger-pixel placements.

\begin{algorithm}[htbp]
    \caption{Computation for Trigger-pixel Placements}
    \label{alg:triloc}
    \begin{algorithmic}[1]
    \renewcommand{\algorithmicrequire}{\textbf{Input:}}
    \renewcommand{\algorithmicensure}{\textbf{Output:}}
    \REQUIRE $W_g$, $D$, $y_t$, $\mathit{tri}_{\mathit{size}}$\\
    \ENSURE  $E_t$

    \STATE $\forall x \in D: y_x \leftarrow W_g(x)$. 
    \STATE $\mathcal{L} \leftarrow \frac{1}{\mid D \mid} \sum_{x \in D} (y_x - y_t)^2 $. 
    \STATE $  \forall x \in D:\delta_{x} \leftarrow \frac{\partial \mathcal{L}}{\partial x}$.
    \STATE $\delta \leftarrow abs(\sum_{x \in D} \delta_{x}) $. 
    \STATE $\delta_{f} \leftarrow $ flatten $\delta$ into a one-dimensional array.
    \STATE $S \leftarrow \text{{argsort}}(\delta_f)$.\COMMENT{\small Store the sorted indices (descending sort)}
    \STATE $E_t \leftarrow S[:\mathit{tri}_{\mathit{size}}]$. \COMMENT{\small Top $\mathit{tri}_{\mathit{size}}$ indices are trigger placements}
    \STATE $E_t \leftarrow$ transform from one-dimensional indices to indices for $x\in D$.

    \RETURN $E_t$
    \end{algorithmic}
\end{algorithm}

Algorithm \ref*{alg:triloc} takes several inputs, including the global
model $W_g$, the trigger training dataset $D$, the target label $y_t$,
and a parameter $\mathit{tri}_{\mathit{size}}$ that specifies the
trigger size. The trigger size $\mathit{tri}_{\mathit{size}}$
determines the number of pixel locations we will choose. The output of
the Algorithm~\ref*{alg:triloc} is the trigger-pixel placement information
denoted as $E_t$.

Starting from line~1 and line~2, we first calculate the loss of the
global model $W_g$ in predicting clean images in dataset $D$ as the
target label $y_t$, where we show Mean Square Error (MSE) as an
example loss function. Next, we compute the gradient of the loss with
respect to each pixel in each image and store the values of gradients
in each image $x$ in $\delta_x$ (line~3). After summing up $\delta_x$
per pixel and take the absolute value of the results, we obtain an
absolute gradient value matrix with the same shape as an individual
image in dataset $D$ (line~4).  To better describe how we sort
elements in $\delta$ by their values, we first flatten $\delta$ into a
one-dimensional array $\delta_f$ (line~5), and then sort elements in
this array in descending order and store the sorted indices in an
array $S$ (line~6). The top $\mathit{tri}_{\mathit{size}}$ number of
indices are the trigger-pixel placements of interest, but before returning
these indices, we transform them from indices for a
one-dimensional array to indices for a matrix of an image's shape in
dataset $D$ (line~7,line~8).

{\bf Optimize trigger-pixel values $V_t$.} In
Algorithm~\ref*{alg:trival}, we optimize the values of the trigger
pixels defined in $E_t$ using a learning-based approach.

\begin{algorithm}[htbp]
    \caption{Optimization for Trigger-pixel Values}
    \label{alg:trival}
    \begin{algorithmic}[1]
    \renewcommand{\algorithmicrequire}{\textbf{Input:}}
    \renewcommand{\algorithmicensure}{\textbf{Output:}}
    \REQUIRE $E_t$, $W_g$, $D$, $y_t$, $n_{\mathit{iter}}$, $\gamma$ \\
    \ENSURE  $V_t$
   \FOR {$\mathit{iteration} \leftarrow 1$ to $n_{\mathit{iter}}$}
       \STATE $D' \leftarrow D$.
       \IF {$\mathit{iteration} = 1$}
        \STATE $V_t \leftarrow \frac{1}{\mid D' \mid}\sum_{x \in D'} x$.
       \ELSIF {$\mathit{iteration} > 1$}
       \STATE $\forall x \in D': x[E_t] \leftarrow V_t[E_t]$.
       \ENDIF
       
       \STATE $\forall x \in D': y_x \leftarrow W_g(x)$.
       \STATE $\mathcal{L} \leftarrow \frac{1}{\mid D' \mid} \sum_{x \in D'} (y_x - y_t)^2$.
       \STATE $\forall x \in D':\delta_{x} \leftarrow \frac{\partial \mathcal{L}}{\partial x}$.
       \STATE $\delta \leftarrow \sum_{x \in D'} \delta_{x} $.
       \STATE $V_t[E_t] \leftarrow (V_t - \gamma\cdot\delta)[E_t]$.
   \ENDFOR
    \RETURN $V_t$
    \end{algorithmic}
\end{algorithm}

Algorithm \ref*{alg:trival} requires the following inputs: the trigger-pixel placements $E_t$, the global model $W_g$, the trigger training dataset
$D$, and the target label $y_t$. Additionally, it uses two training
parameters: the number of training iterations $n_{\mathit{iter}}$ and
the learning rate $\gamma$. The output produced by
Algorithm~\ref*{alg:trival} is the trigger-pixel value information denoted
as $V_t$.

The first step of each iteration is making a copy dataset $D'$ of $D$
(line~2) so that the optimized trigger of each iteration can always be
embedded into clean data. In the first iteration, we initialize the
trigger-pixel value matrix $V_t$ by taking the mean value of all images in
dataset $D'$ along each pixel location (line~4). Then, we calculate
the loss of the global model $W_g$ in predicting images from $D'$ as
the target label $y_t$ (line~8,~9).  Next, we compute the gradients of
the loss with respect to each pixel in each image in dataset $D'$ and
store the values of gradients in each image x in $\delta_x$
(line~10). The gradient matrix $\delta$ is obtained by summing up
$\delta_{x}$ along each pixel location (line~11) (but not need to take
the absolute value as Algorithm \ref{alg:triloc}). After that, we use
the gradient descent technique with $\gamma$ as the learning rate to
only update the values of pixels within the trigger-pixel placements $E_t$
(line~12) and assign those new values to the trigger value matrix
$V_t$. For all iterations after the initial one, we consistently
replace pixels within the trigger-pixel placements $E_t$ of each image with
their corresponding values in the trigger value matrix $V_t$
(line~6). The steps of line 6 and line 12 ensure that the only
variables influencing the loss result are the pixels specified by
$E_t$.

\subsection{Poisoning Malicious Clients' Training Data}

The last step of our attack is to poison malicious clients' local
training data using the optimized trigger $\tau = (E_t, V_t)$ and its
target label $y_t$ by a certain data poison rate. The data poison rate
can be specified on a scale from 0 to 1, while smaller data poison
rate induces stealthier model updates, making them more difficult for
defenses to detect and filter. In the following, we set the data
poison rate to 0.5 for all experiments.  

\section{Theoretical Analysis}\label{sec:thry}

In this section, we delve into the reasons behind \ourmodel{}'s
ability to successfully bypass state-of-the-art defenses, and analyze
the improvements of an optimized trigger generated by our algorithms
in assisting backdoor attacks, compared to a fixed trigger.

We use a linear regression model to explain the intuition of this
work.  Consider a regression problem to model the relationship between
a data sample and its predicted values. We define
$x\in\mathcal{D}^{1\times n}$, where $\mathcal{D}$ is a convex subset
of $\mathbb{R}$ as a data sample, and the vector
$\hat{y}\in\mathbb{R}^{1\times m}$ as its target values.  The model
$\beta\in\mathbb{R}^{n\times m}$ that makes $ x \beta = \hat{y}$ is
what we want to solve.

For any given data $x$, a backdoor attack is aiming to make the model
$\beta$ fit both the benign data point $(x,\hat{y})$ and the
corresponding malicious data point $(x_t,y_t)$. We use
$y_t\in\mathbb{R}^{1\times m}$ to represent the backdoor target values and specify that $y_t \neq \hat{y}$. $x_t\in\mathcal{D}^{1\times
  n}$ is the data $x$ embedded with a trigger $\tau$ by the following
operation.
\begin{equation}\label{eq:operator}
    x_t = x (I_n - E_t) + V_t E_t,
\end{equation}
where $V_t\in\mathcal{D}^{1\times n}$ is a vector storing the trigger
$\tau$'s value information, and $E_t\in\{0,1\}^{n\times n}$ is a
matrix identifying the trigger $\tau$'s location information. $E_t$ specifies the location and shape of the trigger, defined as $E_t = \mathit{diag}(d_1, d_2,..., d_n), d_i \in \{0,1\}$, where $\sum_{i=1}^{n}d_i = k$. Here, $k$ defines the number of entries in the original $x$ that we intend to alter. The abbreviation $\mathit{diag}(\cdot)$ stands for a diagonal matrix whose diagonal values are specified by its arguments. $I_n$ is an $n \times n$ identity matrix.

\begin{definition}{\textbf{(Benign Loss and Benign Objective)}}
Let $x\in\mathcal{D}^{1\times n}$ be a benign data sample,
$\hat{y}\in\mathbb{R}^{1\times m}$ be the predicted value of $x$, and
$\beta\in\mathbb{R}^{n\times m}$ be the prediction model. The loss to
evaluate the prediction accuracy of $\beta$ on the benign regression
is
    \begin{equation}\label{eq:bnloss}
        L(x,\hat{y})=\parallel x\beta - \hat{y} \parallel^2_2.
    \end{equation}
    The optimization objective to solve for $\beta$ for this benign task is 
    \begin{equation}\label{eq:bnobj}
        \min_{\beta}\quad L(x,\hat{y}).
    \end{equation}

\end{definition}

\begin{definition}{\textbf{(Backdoor Loss and Backdoor Objective )}}
Let $x_t$ be a backdoored data sample embedded with a trigger
$\tau(V_t, E_t, y_t)$. Let
$\beta\in\mathbb{R}^{n\times m}$ be the prediction model. The loss to
evaluate the prediction accuracy of $\beta$ on the backdoor regression
is
    \begin{equation}\label{eq:bdloss}
        L(x_t,y_t)= \parallel x_t\beta - y_t \parallel^2_2.
    \end{equation}
The optimization objective to solve for $\beta$ for the backdoor task
that considers both benign data and backdoor data is
    \begin{equation}\label{eq:bdobj}
        \min_{\beta} \quad (1-\alpha) L(x,\hat{y}) + \alpha L(x_t,y_t), 0 \leq \alpha \leq 1.
    \end{equation}

\end{definition}

The FL global model learns backdoor information only when it integrates malicious clients' model updates that were trained for the backdoor objective. Due to the implementation of robust aggregation, backdoor attackers have to ensure their model updates have limited divergence from those trained on benign data to avoid being filtered out by defense techniques. We term this intention as the concealment objective.

To formulate the above problem, we use gradients of optimizing the
benign objective ($G_{\mathit{bn}}$) and gradients of optimizing the
backdoor objective ($G_{\mathit{bd}}$) with respect to a same model
$\beta$ to represent model updates of a benign client and a malicious
client respectively.  We then use cosine similarity as a metric to
evaluate the difference between $G_{\mathit{bn}}$ and
$G_{\mathit{bd}}$, since it is a widely used metric in the
state-of-the-art
defenses~\cite{cao2020fltrust,nguyen2022flame,flair,foolsgold} to
filter malicious model updates.

$G_{\mathit{bn}}$ and $G_{\mathit{bd}}$ are computed by
\begin{subequations}
    \begin{align}
        G_{\mathit{bn}} &=  \frac{\partial L(x,\hat{y})}{\partial \beta},  \label{eq:G_bn}\\
        G_{\mathit{bd}} &=  \frac{\partial ((1-\alpha)L(x,\hat{y})+\alpha L(x_t,y_t))} {\partial \beta}.  \label{eq:G_bd}
    \end{align}
\end{subequations}

The concealment objective is 
\begin{equation} \label{eq:attackgoal}
    \max \quad CosSim(G_{\mathit{bn}}, G_{\mathit{bd}}).
\end{equation}

The optimization objective used in \ourmodel{} attack is
\begin{equation} \label{eq:optbackdoor}
    \min_{V_t,E_t} \quad \parallel (x (I_n - E_t) + V_t E_t)\beta - y_t \parallel^2_2.
\end{equation}

\begin{proposition}\label{thrm: cossim}
Given a model $\beta$ and a data sample $x$ with its benign predicted
value $\hat{y}$ and a backdoor predicted value $y_t$, the optimization
of objective \eqnref{eq:optbackdoor} is a guarantee of the optimization
of objective \eqnref{eq:attackgoal}.
\end{proposition}

\begin{proof}
    See proof in Appendices~\ref{proof: cossim}.
\end{proof}


Proposition~\ref{thrm: cossim} offers a theoretical justification for \ourmodel{}'s ability to prevent malicious clients' model updates from being detected by a commonly used metric considered in state-of-the-art defenses. In Proposition \ref{thrm:2} and Proposition \ref{thrm:3}, we demonstrate that an optimized trigger ($\hat{\tau}$) generated by learning the parameters of a given model $\beta$ is more conducive to achieving the concealment objective compared to a trigger ($\tau_f$) with fixed value, shape, and location.

\begin{proposition}\label{thrm:3}
For any fixed trigger $\tau_f(V_t, E_t, y_t)$ with specified trigger
value $V_t$, trigger location $E_t$, and predicted value $y_t$, there
exists a backdoor trigger $\hat{\tau}(\hat{V_t}, \hat{E_t}, y_t)$ that
has the same $y_t$, but optimizes the $V_t$ and
$E_t$ with respect to a model $\beta$, which can result in a
smaller or equal backdoor loss on model $\beta$ compared to $\tau_f$.
\end{proposition}


\begin{proof}
   See proof in Appendices~\ref{proof:3}.
\end{proof}

\pgfplotsset{width=\linewidth, height=0.5\linewidth, compat=1.15}
\pgfplotsset{grid style={dotted,white!50!black}}
\renewcommand{\arraystretch}{1}

\section{Experiments}\label{sec:exp}

\subsection{Experimental Setup}

\myparatight{Datasets and global models} 
We evaluated \ourmodel{} on four classification datasets with non-IID
data distributions: Fashion MNIST, FEMNIST, CIFAR10, and Tiny ImageNet. Table~\ref{tbl:datadisc} summarizes their basic information
and models we used on each dataset.

\begin{table}[!htbp]
  \renewcommand{\arraystretch}{1.1}
    \caption{Dataset description}
    \vspace{-4mm}
    \begin{center}
      \footnotesize
        \begin{tabular*}{\linewidth}{M{1.2cm}|M{0.6cm}|M{0.55cm}|M{1.1cm}|M{1.2cm}|M{1cm}}
            \hline
        Dataset & \#class& \#img & img size & Model & \#params \\
        \hline
        Fashion MNIST &10& 70k & $28\times28$ grayscale & 2 conv 3 fc & $\sim$1.5M\\
        \hline
        FEMNIST & 62& 33k & $28\times28$ grayscale  & 2 conv 2 fc & $\sim$6.6M\\
        \hline
        CIFAR10 &10& 60k & $32\times32$ color  & ResNet18 & $\sim$11M\\
        \hline 
        Tiny ImageNet &200& 100k & $64\times64$ color  & VGG11 & $\sim$35M\\
        \hline
        \end{tabular*}
    \end{center}
    \label{tbl:datadisc}
    \vspace{-6mm}
  \end{table}

\myparatight{Comparisons}
As \ourmodel{} is exclusively a data-poisoning attack, we compared it
with existing attacks where all the non-data-poisoning components were
removed. To be specific, we only implemented the trigger embedding part
introduced in existing attacks, while disregarding any model-poisoning
techniques such as objective modification, alterations to training
hyperparameters, or scaling up malicious model updates. 

We compared \ourmodel{} with three existing attacks as described below. 

\begin{itemize}[leftmargin=*]
  \vspace{-2mm}
  \setlength{\itemsep}{0pt}
  \setlength{\parskip}{0pt}
  \setlength{\parsep}{0pt}
\item {\bf Fixed Trigger (\textbf{FT})}. Following recent research on
backdoor attacks on FL~\cite{alittleisenough, Xie2020DBA:,
cao2020fltrust, howtobackdoor}, pixel-pattern triggers are typical
backdoors applied in image classification applications.  A
pixel-pattern trigger is a defined arrangement of pixels with specific
values and shape, placed at a particular
location within images. We used a global pixel-pattern
trigger with fixed features (values, shape, and placement) for all experiments in this attack category.

\item {\bf Distributed Fixed Trigger (\textbf{DFT})}. Inheriting the
definition of the pixel-pattern trigger, DBA~\cite{Xie2020DBA:} slices
a global pixel-pattern trigger into several parts and distributes
them among different malicious FL clients for data poisoning. The Attack Success Rate for this attack category is evaluated based on the global pixel-pattern trigger.

\item {\bf A3FL Trigger}. A state-of-the-art attack in FL,
A3FL~\cite{zhang2024a3fl}, proposed adversarially optimizing the
trigger's value using a local model that continuously unlearns the
optimized trigger information. The shape and placement of the A3FL
trigger stay fixed during optimization. We compare their methods on
CIFAR10 dataset as it is the only available configuration in their
open-source project.
\vspace{-2mm}
\end{itemize}

The visualization of various trigger types are demonstrated in
Figures~\ref{fig:poisoned_ft_ti}, \ref{fig:poisoned_dft_ti},
\ref{fig:poisoned_opt_ti}, and \ref{fig:poisoned_A3FL}.

\myparatight{Defenses}
We evaluated backdoor attacks in FL systems employing different
state-of-the-art defense strategies against backdoor attacks.  We
selected defense baselines based on two criteria: 1) Defenses provided
accessible proof-of-concept codes to ensure accurate implementation of
their proposed ideas; 2) Defenses either claimed or were proven by
existing research to have defense effectiveness against backdoor
attacks in FL. 

We presented the evaluation results of 10 defense strategies that rely solely on server-side execution in Section~\ref{sec:exp}. Detailed descriptions of these defenses were presented in Appendices~\ref{defensedescription}. The evaluation results of defenses requiring client-side execution, Flip~\cite{zhang2022flip} and FRL~\cite{everyvotecounts}, were demonstrated in Appendices~\ref{flip} and Appendices~\ref{FRL} due to space limitations.

\myparatight{Evaluation metrics}
We considered three metrics to evaluate the effectiveness and
stealthiness of backdoor attacks when confronted with different
defense strategies.

\textbf{Final Attack Success Rate (Final $\mathit{ASR}$)}. This metric
quantifies the proportion of backdoored test images that were misclassified
as the target label by the global model at the end of training. In order to reduce the
testing error caused by noise on data or model so as to maintain the
fairness of comparison, we tested $\mathit{ASR}$ on the global models of
the last five rounds and took their mean value as the Final
$\mathit{ASR}$.

\textbf{Average Attack Success Rate (\textbf{Avg
$\mathit{ASR}$})}. We initiate the attack from the first round of FL
training. Since the attack cycle of \ourmodel{} spans just a single
round, we introduced Avg $\mathit{ASR}$ to assess the average attack
effectiveness across all rounds during the FL process. To
evaluate $\mathit{ASR}$ for an individual round, we poison test data
with the optimized trigger $\tau^{(i)}$ which is generated using the current-round global
model ${W_g}^{(i)}$ and test its
$\mathit{ASR}$ on the next-round global model ${W_g}^{(i+1)}$.  We
took the average of the $\mathit{ASR}$ of all rounds in the FL process
as the Avg $ASR$. The implication of a high Avg $\mathit{ASR}$ of an
attack is that this attack had consistently significant effectiveness
during the whole FL process, ensuring a high Final $\mathit{ASR}$ no
matter when the FL process ended.

\textbf{Main-task Accuracy (\textbf{$\mathit{MA}$})}. We evaluate this metric by testing the accuracy of a global model on its clean main-task test dataset. A backdoor attack is seen
to be stealthy if its victim model does not show a noticeable reduction
in $\mathit{MA}$ compared to the benign model.

\myparatight{FL configurations} The FEMNIST
dataset~\cite{caldas2018leaf} provides each client's local training data with a
naturally non-IID guarantee.  For Fashion MNIST, CIFAR10, and Tiny ImageNet
datasets, we distributed training data to FL clients using the same
method introduced by FLTrust~\cite{cao2020fltrust}, where we set the
non-IID bias to be 0.5.

For all datasets training experiments, we used SGD
optimization with CrossEntropy loss. In the experiments on Tiny ImageNet, we set the mini-batch size to 64, while for the other datasets, we set it to 256. Each
FL client trained a global model for $n_{\mathit{epoch}} = 5$ local epochs with its local data in one global round.

For training Fashion MNIST and FEMNIST datasets, we used a
static local learning rate ($\mathit{lr}$) of 0.01. For training the larger and more complicated datasets such as CIFAR10 and Tiny ImageNet, we applied the learning rate schedule technique following the instructions in the related machine learning works~\cite{he2016deep, simonyan2014very} to boost DNN models' performance.

\myparatight{Attack configurations} In our
algorithm \ref{alg:trival} for training a trigger value, we set the
number of training iterations to $n_{\mathit{iter}} = 10$, which
proved sufficient for obtaining the optimal trigger values. The
learning rate $\gamma$ started from 5 and was halved when the training
loss increased compared to the previous iteration.

Table~\ref*{tbl:defaultsetting} shows the default settings of \ourmodel{} attack for experiments. In particular, we consider the following attributes that are critical for attack effectiveness.

\begin{table}[!htbp]
  \renewcommand{\arraystretch}{1.1}
  \caption{Default Settings} \vspace{-4mm} 
    \begin{center} 
      \footnotesize
      \begin{tabular*}{\linewidth}{M{1.4cm}|M{1.1cm}|M{1.3cm}|M{1.1cm}|M{1.2cm}} 
        \hline
    &\scriptsize Fashion MNIST & \scriptsize FEMNIST & \scriptsize CIFAR10
    &\scriptsize Tiny ImageNet\\ 
    \hline 
    Trigger Size & 64 & 25 & 25 & 64 \\ 
    \hline 
    Round &300 &200 &150 &100\\ 
    \hline 
    Number of Clients&100 &100 & 50& 50\\ 
    \hline
    MCR&\multicolumn{4}{c}{0.05} \\ 
    \hline 
    \scriptsize Local Data Poison\-Rate &\multicolumn{4}{c}{0.5} \\ \hline
    \end{tabular*}
    \end{center}
    \label{tbl:defaultsetting}
    \vspace{-4mm}
    \end{table}

\textbf{Trigger size.} We defined trigger size for different
datasets according to the following three criteria. First, a trigger
of the defined trigger size should not be able to cover important
details of any images and lead humans to misidentify the images from
their original labels. To show that we follow this criteria, we
demonstrated poisoned images from different datasets that are embedded
with \ourmodel{} triggers, as shown in
Figure~\ref{fig:poisoned_img_fe_cf} and ~\ref{fig:poisoned_opt}. Second, on basis of the first
criterion, we adjust trigger size to match the image size and the
feature size of different datasets. Specifically, if a dataset
contains images with high resolution (large image size), then a large
trigger size is needed to effectively match it (Tiny ImageNet vs.\
CIFAR10).  If images in a dataset contain large visual elements or patterns, then a large
trigger size is needed to effectively match it (Fashion MNIST vs.\
FEMNIST). Third, we found that when using models with deep model
architectures or having large number of parameters, a small trigger
size is sufficient for conducting \ourmodel{} attack (CIFAR10
vs.\ Fashion MNIST).

\textbf{Round.} We determine the number of training rounds for each
dataset by measuring the convergence time on an FL pipeline using
FedAvg as the aggregation rule. Convergence is considered achieved
when the test accuracy on the main task stabilizes within a range of
0.5 percentage points over a period of five consecutive rounds of training.

\textbf{Number of clients.} The number of clients varies across
different datasets due to a balance between our available
computational resources and the size of the datasets/models.  All
clients participate in the aggregation for each round of FL training.

\textbf{MCR.} Malicious Client Ratio (MCR) is a parameter defining the
proportion of compromised clients compared to the total number of
clients in each-round aggregation. We consider $5\%$ as the default
MCR (for FL systems having 50 clients, 2 of them are malicious
clients), which is smaller than the state-of-the-art
attacks~\cite{zhang2024a3fl,CerP,fang2023vulnerability} that require
at least $10\%$ of clients to behave maliciously during aggregation.

\textbf{Local data poison rate.} It indicates the proportion of data manipulated by a backdoor attack relative to the total data available on each malicious client.

\myparatight{Experiment environment and code} We conducted all the
experiments on a platform with multiple NVIDIA Quadro RTX 6000 Graphic Cards
having 24 GB GPU memory in each chip and an Intel(R) Xeon(R) Gold 6230
CPU @2.10GHz having 384 GB CPU memory. We implemented all the
algorithms using the PyTorch framework. We will open-source this project after its publication.

\subsection{Experimental Results}


\subsubsection{Representative Results}
\begin {figure}[!th]
    \centering
    \vspace{-3mm}
    \begin{subfigure}[h]{1.04\linewidth}
        \begin{tikzpicture}[
            every axis/.style={ 
            ybar,
            ymin=0,ymax=105,
            x tick label style={rotate=0},
            symbolic x coords={
            FedAvg, Median, {\footnotesize Trimmed Mean},
            RobustLR, RFA, FLAIR, FLCert,
            FLAME, FoolsGold, Multi-Krum
            },
            ylabel = {$ASR$},
            bar shift=0pt,
            x tick label style={rotate=30,anchor=east, font=\small},
            y tick label style = {font=\small},
            label style={ font=\small},
            xtick=data,
            bar width=5pt,
            legend style={at={(0.97,1.5)},legend columns=3, font=\small},
            ymajorgrids=true},
        ]
        \begin{axis}[bar shift=-5pt, hide axis]
        \addplot+[only marks,mark=triangle*,mark options={fill=black,draw=black}, xshift=-5pt, legend image post style={xshift=5pt}] coordinates
        {(Median,61.6694) (FedAvg,69.1043) ({\footnotesize Trimmed Mean},55.9902) (RobustLR, 62.8) (RFA, 62.0) (FLAIR,51.6) (FLCert,57.8812) (FLAME,43.42896667) (FoolsGold,68.48426667) (Multi-Krum,63.6)};\label{bar0}
        \addplot+[black, fill=black!30!white, mark=no] coordinates
        {(Median,97.78) (FedAvg,97.716) ({\footnotesize Trimmed Mean},94.4) (RobustLR, 99.2)(RFA,97.7) (FLAIR,85.1) (FLCert,95.204) (FLAME,71.09) (FoolsGold,98.854) (Multi-Krum,99.9)};\label{bar1}
      
        \end{axis}
        \begin{axis} [hide axis]
        \addplot+[only marks,mark= triangle*,mark options={fill=black!50!white, draw=black}] coordinates
        {(Median,11.0421) (FedAvg,15.12823333) ({\footnotesize Trimmed Mean},11.035)(RobustLR, 28.5) (RFA,12.2) (FLAIR,15.8) (FLCert,10.713) (FLAME,10.63686667) (FoolsGold,14.21516667) (Multi-Krum,10.7)};\label{bar2}
        \addplot+[black, fill=black!10!white, pattern=north east lines, mark=no] coordinates
        {(Median,10.318) (FedAvg,33.98) ({\footnotesize Trimmed Mean},10.176)(RobustLR, 85.4) (RFA,12.2) (FLAIR,26.9) (FLCert,10.266) (FLAME,9.596) (FoolsGold,29.488) (Multi-Krum,9.8)};\label{bar3}
      
        \end{axis}
        \begin{axis}[bar shift=5pt]
            \addlegendimage{/pgfplots/refstyle=bar1}
            \addlegendentry{Ours  (Final)}
            \addlegendimage{/pgfplots/refstyle=bar3}
            \addlegendentry{FT (Final)}
            \addlegendimage{black,fill=white, mark=no}
            \addlegendentry{DFT (Final)}
            \addlegendimage{/pgfplots/refstyle=bar0}
            \addlegendentry{Ours  (Avg)}
            \addlegendimage{/pgfplots/refstyle=bar2}
            \addlegendentry{FT  (Avg)}
            \addlegendimage{only marks,mark=triangle*,mark options={fill=white,draw=black}}
            \addlegendentry{DFT (Avg)}
        \addplot+[only marks,mark= triangle*,mark options={fill=white,draw=black}, xshift=5pt, legend image post style={xshift=-5pt}] coordinates
        {(Median,10.7053) (FedAvg,10.9564) ({\footnotesize Trimmed Mean},10.64473333)(RobustLR, 11.4) (RFA,11.0) (FLAIR,11.2) (FLCert,10.12906667)  (FLAME,11.17003333) (FoolsGold,10.92563333) (Multi-Krum,10.9)};
        \addplot+[black,fill=white, mark=no] coordinates
        {(Median,10.11) (FedAvg,10.478) ({\footnotesize Trimmed Mean},10.164)(RobustLR, 10.8) (RFA,10.3) (FLAIR,11.2) (FLCert,10.064) (FLAME,9.948) (FoolsGold,10.552) (Multi-Krum,10.0)};
      
        \end{axis}
        \end{tikzpicture}
        \vspace{-5mm}
      \caption{Fashion MNIST}
      \label{fig:fashionmnist_MIR0.05}

      \end {subfigure}

      \begin{subfigure}[h]{1.04\linewidth}
        \begin{tikzpicture}[
        every axis/.style={ 
        ybar,
        ymin=0,ymax=105,
        x tick label style={rotate=0},
        symbolic x coords={
          FedAvg, Median, {\footnotesize Trimmed Mean},
          RobustLR, RFA, FLAIR, FLCert,
          FLAME, FoolsGold, Multi-Krum
        },
        ylabel = {$ASR$},
        bar shift=0pt,
        x tick label style={rotate=30,anchor=east, font=\small},
        y tick label style = {font=\small},
        label style={ font=\small},
        xtick=data,
        bar width=5pt,
        legend style={at={(0.9,1.5)},legend columns=3, font=\small},
        ymajorgrids=true
        },
        ]
        \begin{axis}[bar shift=-5pt, hide axis]
        \addplot+[only marks,mark=triangle*,mark options={fill=black,draw=black}, xshift=-5pt, legend image post style={xshift=5pt}] coordinates
        {(Median,81.17099761) (FedAvg,92.85364397) ({\footnotesize Trimmed Mean},84.28838112)(RobustLR, 93) (RFA, 95.9 ) (FLAIR,72.7) (FLCert,86.70) (FLAME,86.14829749) (FoolsGold,95.20564516) (Multi-Krum,92.0)};\label{bar0}
        \addplot+[black, fill=black!30!white, mark=no] coordinates
        {(Median,95.35842294) (FedAvg,99.74313023) ({\footnotesize Trimmed Mean},95.22102748)(RobustLR, 99) (RFA, 98.3) (FLAIR,88.7) (FLCert,97.09) (FLAME,99.22939068) (FoolsGold,99.55197133) (Multi-Krum,99.7)};\label{bar1}
      
        \end{axis}
        \begin{axis} [hide axis]
        \addplot+[only marks,mark= triangle*,mark options={fill=black!50!white, draw=black}] coordinates
        {(Median,54.02837515) (FedAvg,88.01015532) ({\footnotesize Trimmed Mean},44.01612903)(RobustLR, 80) (RFA, 79.4) (FLAIR,12.9) (FLCert,66.76) (FLAME,82.93817204) (FoolsGold,77.35961768) (Multi-Krum,64.4)};\label{bar2}
        \addplot+[black, fill=black!10!white, pattern=north east lines, mark=no] coordinates
        {(Median,76.23655914) (FedAvg,94.08602151) ({\footnotesize Trimmed Mean},64.27120669)(RobustLR, 96) (RFA, 94.8) (FLAIR,7.5) (FLCert,81.57) (FLAME,92.59856631) (FoolsGold,96.24253286) (Multi-Krum,74.0)};\label{bar3}
      
        \end{axis}
        \begin{axis}[bar shift=5pt]

        \addplot+[only marks,mark= triangle*,mark options={fill=white,draw=black}, xshift=5pt, legend image post style={xshift=-5pt}] coordinates
        {(Median,7.483870968) (FedAvg,14.41129032) ({\footnotesize Trimmed Mean},32.1135006)(RobustLR, 55) (RFA, 44.9) (FLAIR,3.4) (FLCert,12.59543011)  (FLAME,78.96176822) (FoolsGold,4.750448029) (Multi-Krum,62.9)};
        \addplot+[black,fill=white, mark=no] coordinates
        {(Median,11.96535245) (FedAvg,36.23655914) ({\footnotesize Trimmed Mean},55.43010753)(RobustLR, 98) (RFA, 92.0) (FLAIR,3.0) (FLCert,22.06690562) (FLAME,99.95221027) (FoolsGold,7.897252091) (Multi-Krum,95.9)};
      
        \end{axis}
        \end{tikzpicture}
        \vspace{-5mm}
      \caption{FEMNIST}
      \label{fig:femnist_MIR0.05}

      \end {subfigure}

\begin{subfigure}[h]{1.04\linewidth}
    \begin{tikzpicture}[
    every axis/.style={ 
    ybar,
    ymin=0,ymax=105,
    x tick label style={rotate=0},
    symbolic x coords={
      FedAvg, Median, {\footnotesize Trimmed Mean},
      RobustLR, RFA, FLAIR, FLCert,
      FLAME, FoolsGold, Multi-Krum
    },
    ylabel = {$ASR$},
    bar shift=0pt,
    x tick label style={rotate=30,anchor=east, font=\small},
    y tick label style = {font=\small},
    label style={ font=\small},
    xtick=data,
    bar width=5pt,
    ymajorgrids=true
    },
    ]

    \begin{axis}[bar shift=-5pt, hide axis]
    \addplot+[only marks,mark=triangle*,mark options={fill=black,draw=black}, xshift=-5pt, legend image post style={xshift=5pt}] coordinates
    {(Median,96.11) (FedAvg,98.53) ({\footnotesize Trimmed Mean},88.60)(RobustLR, 98.6) (RFA, 98) (FLAIR,50.7) (FLCert,88.32) (FLAME,56.04) (FoolsGold,98.46) (Multi-Krum,98.7)};\label{bar0}
    \addplot+[black, fill=black!30!white, mark=no] coordinates
    {(Median,99.98) (FedAvg,100) ({\footnotesize Trimmed Mean},99.95)(RobustLR, 100.0) (RFA,100) (FLAIR,62.3) (FLCert,99.16) (FLAME,59.812) (FoolsGold,100) (Multi-Krum,100.0)};\label{bar1}

    \end{axis}

    \begin{axis} [hide axis]
    \addplot+[only marks,mark= triangle*,mark options={fill=black!50!white, draw=black}] coordinates
    {(Median,46.64) (FedAvg,88.06) ({\footnotesize Trimmed Mean},59.41)(RobustLR, 94.4) (RFA,81) (FLAIR,13.6) (FLCert,59.52) (FLAME,18.19) (FoolsGold,88.41) (Multi-Krum,98.7)};\label{bar2}
    \addplot+[black, fill=black!10!white, pattern=north east lines, mark=no] coordinates
    {(Median,81.31) (FedAvg,100) ({\footnotesize Trimmed Mean},94.81)(RobustLR, 100.0) (RFA,100) (FLAIR,15.1) (FLCert,92.58) (FLAME,19.00) (FoolsGold,99.99) (Multi-Krum,100)};\label{bar3}

    \end{axis}

    \begin{axis}[bar shift=5pt]
    \addplot+[only marks,mark= triangle*,mark options={fill=white,draw=black}, xshift=5pt, legend image post style={xshift=-5pt}] coordinates
    {(Median,41.76) (FedAvg,50.24) ({\footnotesize Trimmed Mean},22.64)(RobustLR, 87.4) (RFA,55) (FLAIR,10.1) (FLCert,21.34)  (FLAME,14.35) (FoolsGold,52.95) (Multi-Krum,82.8)};

    \addplot+[black,fill=white, mark=no] coordinates
    {(Median,72.01) (FedAvg,92.5) ({\footnotesize Trimmed Mean},38.28)(RobustLR, 100.0) (RFA,98) (FLAIR,9.7) (FLCert,33.72) (FLAME,16.84) (FoolsGold,94.13) (Multi-Krum,99.9)};

    \end{axis}
    \end{tikzpicture}
    \vspace{-5mm}
\caption{CIFAR10}
\label{fig:cifar_MIR0.05}

\end {subfigure}

\begin{subfigure}[h]{1.04\linewidth}
    \begin{tikzpicture}[
        every axis/.style={ 
        ybar,
        ymin=0,ymax=105,
        x tick label style={rotate=0},
        symbolic x coords={
          FedAvg, Median, {\footnotesize Trimmed Mean},
          RobustLR, RFA, FLAIR, FLCert,
          FLAME, FoolsGold, Multi-Krum
        },
        ylabel = {$ASR$},
        bar shift=0pt,
        x tick label style={rotate=30,anchor=east, font=\small},
        y tick label style = {font=\small},
        label style={ font=\small},
        xtick=data,
        bar width=5pt,
        legend style={at={(0.9,1.5)},legend columns=3, font=\small},
        ymajorgrids=true},
    ]
    \begin{axis}[bar shift=-5pt, hide axis]
    \addplot+[only marks,mark=triangle*,mark options={fill=black,draw=black}, xshift=-5pt, legend image post style={xshift=5pt}] coordinates
    {(Median,30.51) (FedAvg,62.19) ({\footnotesize Trimmed Mean},36.41)(RobustLR, 64.09) (RFA, 67.75) (FLAIR,35.27) (FLCert,32.30) (FLAME,40.75) (FoolsGold,63.65) (Multi-Krum,82.10)};\label{bar0}
    \addplot+[black, fill=black!30!white, mark=no] coordinates
    {(Median,90.77) (FedAvg,97.90) ({\footnotesize Trimmed Mean},94.59)(RobustLR, 98.22) (RFA,99.45) (FLAIR,92.89) (FLCert,72.44) (FLAME,91.05) (FoolsGold,98.92) (Multi-Krum,99.93)};\label{bar1}
  
    \end{axis}
    \begin{axis} [hide axis]
    \addplot+[only marks,mark= triangle*,mark options={fill=black!50!white, draw=black}] coordinates
    {(Median,17.05) (FedAvg,45.35) ({\footnotesize Trimmed Mean},29.26)(RobustLR, 53.47) (RFA,54.67) (FLAIR,17.07) (FLCert,23.70) (FLAME,40.29) (FoolsGold,36.43) (Multi-Krum,95.46)};\label{bar2}
    \addplot+[black, fill=black!10!white, pattern=north east lines, mark=no] coordinates
    {(Median,60.61) (FedAvg,94.11) ({\footnotesize Trimmed Mean},84.26)(RobustLR, 98.26) (RFA,99.72) (FLAIR,78.17) (FLCert,58.52) (FLAME,70.92) (FoolsGold,89.23) (Multi-Krum,100.00)};\label{bar3}
  
    \end{axis}
    \begin{axis}[bar shift=5pt]
    \addplot+[only marks,mark= triangle*,mark options={fill=white,draw=black}, xshift=5pt, legend image post style={xshift=-5pt}] coordinates
    {(Median,12.92) (FedAvg,27.97) ({\footnotesize Trimmed Mean},12.56)(RobustLR, 30.63) (RFA,31.70) (FLAIR,31.93) (FLCert,9.88)  (FLAME,36.69) (FoolsGold,30.78) (Multi-Krum,85.73)};
    \addplot+[black,fill=white, mark=no] coordinates
    {(Median,46.80) (FedAvg,69.91) ({\footnotesize Trimmed Mean},43.58)(RobustLR, 69.38) (RFA,83.96) (FLAIR,90.58) (FLCert,35.21) (FLAME,80.44) (FoolsGold,68.32) (Multi-Krum,99.49)};
  
    \end{axis}
    \end{tikzpicture}
    \vspace{-4mm}
\caption{Tiny ImageNet}
\label{fig:tiny_MIR0.05}
\vspace{-2mm}
\end {subfigure}
\caption{Representative results on four different datasets are provided. The attack settings correspond to the default settings outlined in Table ~\ref{tbl:defaultsetting}.}
\label{fig:main_ASR}
\vspace{-6mm}
\end {figure}
In this section, we presented the performance of \ourmodel{} attack
under 10 defense methods and compared our results with two
widely-used data-poisoning attacks.

The effectiveness of an attack is measured using the $\mathit{ASR}$
metric, as shown in Figure~\ref{fig:main_ASR}. Results indicate that
the \ourmodel{} attack consistently achieves a final $\mathit{ASR}$
exceeding 50\% across all considered defense methods, regardless of
the dataset's characteristics such as imgae size and number of
images. Additionally, the \ourmodel{} attack also exhibits a
considerable average $\mathit{ASR}$ in each attack practice,
indicating its malicious effect on each-round global model. The
stealthiness of an attack is assessed using the $\mathit{MA}$ metric,
as indicated in Table~\ref{tbl:main_MA}. We established a baseline
$\mathit{MA}$ for each defense method on every dataset by measuring
the final $\mathit{MA}$ achieved in an attack-free FL training session
employing the respective defense. Upon comparing the baseline
$\mathit{MA}$ of various defenses to that of FedAvg, we observed that
certain defenses, such as Multi-Krum on most of datasets and FLAME on Tiny
ImageNet, failed to achieve similar convergence performance as FedAvg
under the same training conditions. Defenses with deficient baseline
$\mathit{MA}$ are less likely to be adopted in practice.  The results
presented in Table~\ref{tbl:main_MA} indicate that the \ourmodel{}
attack successfully maintains the $\mathit{MA}$ of victim global
models within a $\pm2$ percentage-point difference range compared to the corresponding baseline $\mathit{MA}$ values.

In comparison to FT and DFT attacks, the \ourmodel{} attack demonstrates superior attack effectiveness in compromising existing defenses. As illustrated in Figure~\ref{fig:main_ASR}, the \ourmodel{} attack consistently demonstrates a higher Final ASR compared to FT and DFT attacks and also achieves a significantly better Average ASR.


\subsubsection{Comparison to A3FL Trigger} 

\begin{table}[!htbp]
  \renewcommand{\arraystretch}{1.1}
  \footnotesize\addtolength{\tabcolsep}{-2pt}
  \begin{center}
    \begin{tabular}{|c|cc|cc|cc|}
      \hline
       &\multicolumn{2}{c|}{\textbf{Final $\mathit{ASR}$}} &\multicolumn{2}{c|}{\textbf{Average $\mathit{ASR}$}} & \multicolumn{2}{c|}{\textbf{MA}} \\
      \hline
      & \footnotesize Ours &\footnotesize A3FL &\footnotesize Ours &\footnotesize A3FL &\footnotesize Ours &\footnotesize A3FL\\
      \hline
      FedAvg &\textbf{100} &48.9 &\textbf{98.5} &38.1 &70.7 &70.6 \\
      \hline
      Median &\textbf{100} &32.9 &\textbf{96.1} &24.0 &69.1 &69.1 \\
      \hline
      \footnotesize Trimmed Mean &\textbf{100} &35.0 &\textbf{88.6} &23.5 &70.4 &69.9 \\
      \hline
      RobustLR &\textbf{100} &46.2 &\textbf{98.6} &40.7 &70.1 &71.2 \\
      \hline
      RFA &\textbf{100} &24.7 &\textbf{97.8} &23.8 &70.7 &70.2 \\
      \hline
      FLAIR &\textbf{62.3} &13.2 &\textbf{50.7} &12.5 &70.6 &70.7 \\
      \hline
      FLCert &\textbf{99.2} &39.0 &\textbf{88.3} &28.4 &70.0 &69.9 \\
      \hline
      FLAME &\textbf{59.8} &13.7 &\textbf{56.0} &32.1 &70.3 &70.1 \\
      \hline
      FoolsGold &\textbf{100} &46.9 &\textbf{98.5} &38.0 &71.0 &70.8 \\
      \hline
      Multi-Krum &\textbf{100} &33.4 &\textbf{98.7} &29.5 &63.0 &62.8 \\
      \hline
    \end{tabular}
  \end{center}
  \vspace{-5mm}
  \caption{Comparison results with A3FL attack on CIFAR10.}
  \label{tbl:A3FL}
\end{table}

In this section, we compared the performance of \ourmodel{} attack with the
A3FL~\cite{zhang2024a3fl} attack. We implemented the A3FL attack by
faithfully replicating the attacker's actions as designed by A3FL,
with reference to their open-source project. We evaluated the effectiveness of A3FL attack against 10 defense strategies within our FL configurations
and attack settings (refer to Table~\ref{tbl:defaultsetting}).

The results in Table~\ref{tbl:A3FL} demonstrate that our attack
achieved significantly higher $\mathit{ASR}$ values in both the final
and average metrics compared to the A3FL attack. This
suggests that the optimized triggers generated using our algorithms
are more effective in compromising FL global models through data
poisoning compared to those generated using A3FL's
techniques. Additionally, we observed that the $\mathit{ASR}$ results
of A3FL were even worse than those of FT and DFT (as shown in
Figure~\ref{fig:main_ASR}) in our experiment settings. This implies that dynamically changing the
backdoor objective may not enhance the effectiveness of
backdoor attacks compared to maintaining a static backdoor
objective if it can not align to the benign objective effectively.

\vspace{-2mm}
\subsubsection{Analysis of the \ourmodel{} working principles }


In this section, we analyzed the attack effectiveness of each component
of the \ourmodel{} attack's working principles and report evidence
that it effectively conceals malicious clients' model updates, thereby getting them
integrated into the global models through aggregation.

In the $i$-th round, \ourmodel{} generates a trigger $\tau^{(i)} $ by
optimizing its shape, placement and values to make the global model of
this round ${W_g}^{(i)}$ achieve a maximum $\mathit{ASR}$.  However,
what we were truly interested in is its $\mathit{ASR}$ on the global
model after the $i$-th round aggregation, which is the next-round
global model denoted as ${W_g}^{(i+1)}$. The attack effectiveness of
the trigger $\tau^{(i)}$ on the global model ${W_g}^{(i+1)}$ stems
from two factors:
\begin{enumerate}[leftmargin=*]

\item \textbf{Trigger Optimization}: Trigger optimization using ${W_g}^{(i)}$ results in an improvement of the trigger's $\mathit{ASR}$ on ${W_g}^{(i+1)}$ due to the small difference between ${W_g}^{(i+1)}$ and ${W_g}^{(i)}$.  
\item \textbf{Concealment of Model Updates}: Model updates that were trained on data partially poisoned by $\tau^{(i)}$ exhibit small differences from those were trained on data without poisoning. Therefore, they were aggregated into ${W_g}^{(i+1)}$ and made ${W_g}^{(i+1)}$ incorporate backdoored model parameters. 

\end{enumerate}

In the following, we explain how we designed experiments to study the
impact of each factor, and analyzed the experiment results.

\myparatight{Experiment design} 
To assess the attack effectiveness solely brought by
Trigger Optimization, we eliminated any effects produced by data
poisoning. Specifically, we set all clients in the FL
system to be benign, ensuring that the next-round global model, denoted as
$\widetilde{W}_g^{(i+1)}$, aggregated benign model updates only. In the meantime, we still collected data from a certain number of clients and optimized a trigger $\widetilde{\tau}^{(i)}$ for $\widetilde{W}_g^{(i)}$. Then, we
tested $\widetilde{W}_g^{(i+1)}$ on a testing dataset in which all images are
poisoned with the trigger $\widetilde{\tau}^{(i)}$ to obtain an
$\widetilde{ASR}$. This $\widetilde{ASR}$ evaluates the
 attack effectiveness achieved by the current-round optimized trigger
$\tau^{(i)}$ on the next-round global model $\widetilde{W}_g^{(i+1)}$, which does not contain any model updates learned from backdoor information.

To assess the attack effectiveness brought by Concealment of Model Updates, we introduced malicious clients into the FL system and therefore the global model, denoted as $\ddot{W}_g^{(i+1)}$, was allowed to aggregate model updates submitted by malicious clients.  
In this system, malicious clients partially poisoned their local training data (aligning with default settings in Table~\ref{tbl:defaultsetting}) using the trigger $\ddot{\tau}^{(i)}$ that was optimized for $\ddot{W}_g^{(i)}$, and then conducted their local training.   We tested the $\ddot{W}_g^{(i+1)}$ on the testing dataset that was also poisoned by $\ddot{\tau}^{(i)}$ to obtain an $\ddot{ASR}$. We evaluated the attack
effectiveness of Concealment of Model Updates by measuring the increase in
$\mathit{ASR}$ compared to the previous setting, calculated as ($\ddot{ASR} - \widetilde{ASR}$). This metric reveals
how much the malicious clients' model updates influenced the global model
$\ddot{W}_g^{(i+1)}$ to achieve a higher $\mathit{ASR}$ compared to
$\widetilde{W}_g^{(i+1)}$.

\begin{table}[!t] 
  \renewcommand{\arraystretch}{1}
  \small\addtolength{\tabcolsep}{-3pt}
  \begin{center}
    \begin{tabular}{|cc|cc|cc|cc|}
      \hline
      &  & \multicolumn{2}{M{1.4cm}|}{Fashion MNIST} & \multicolumn{2}{c|}{FEMNIST} & \multicolumn{2}{c|}{CIFAR10} \\
      \hline
      & $\mathit{ASR}$ type & Final & Avg  & Final & Avg & Final & Avg\\
      \hline
      \multirow{2}{*}{FedAvg}  &  $\widetilde{ASR}$  & 58.8&45.1 & 54.0&28.6 & 55.6&50.9  \\
      & $\ddot{ASR}$ & 97.7&69.1 & 99.7&92.9 & 100&98.5  \\
      \hline
      \multirow{2}{*}{Median}  & $\widetilde{ASR}$  & 57.9&38.2 & 18.0&17.5 & 56.6&48.7  \\
      & $\ddot{ASR}$ & 97.8&61.7 & 95.4&81.2 &100&96.1  \\
      \hline
       Trimmed  & $\widetilde{ASR}$  & 31.6&29.7 & 24.2&25.6 & 55.6&40.9 \\
      Mean  & $\ddot{ASR}$ & 94.4&56.0 & 95.2&84.3 & 100&88.6  \\
      \hline
      \multirow{2}{*}{RobustLR}  & $\widetilde{ASR}$  & 70.2&47.2 & 28.8& 27.3&60.1 &47.3  \\
      & $\ddot{ASR}$ &99.2 &62.8 &99.3 &93.0 &100.0 &98.6  \\
      \hline
      \multirow{2}{*}{RFA}  & $\widetilde{ASR}$  & 78.0&46.4 &18.9 &13.4 &57.4 &46.1  \\
      & $\ddot{ASR}$ &97.7 &62.0 &98.3 &95.9 &100.0 &97.8  \\
      \hline
      \multirow{2}{*}{FLAIR}  & $\widetilde{ASR}$  & 42.2&36.2 & 23.0&29.6 & 54.1	&45.9  \\
      & $\ddot{ASR}$ & 85.3&50.1 & 88.7&72.7 & 62.3&50.7  \\
      \hline
      \multirow{2}{*}{FLCert}  & $\widetilde{ASR}$  & 49.6&39.7 & 27.7&34.6 & 48.7&46.7  \\
      & $\ddot{ASR}$ & 95.2&57.9 & 97.1&86.7 & 99.2&88.3 \\
      \hline
      \multirow{2}{*}{FLAME}  & $\widetilde{ASR}$  & 38.0&26.2 & 34.7&35.7 & 28.1&51.0  \\
      & $\ddot{ASR}$ & 71.1&43.4 & 99.2&86.1 & 59.8&56.1 \\
      \hline
      Fools- & $\widetilde{ASR}$  & 54.2&50.3 & 57.0&43.7 & 35.5&35.6  \\
      Gold & $\ddot{ASR}$ &98.9&68.5 &99.6&95.2 & 100&98.5  \\
      \hline
      Multi- & $\widetilde{ASR}$  & 60.6&45.4 & 31.7& 28.7& 49.7&36.1  \\
      Krum & $\ddot{ASR}$ & 99.9&63.6 & 99.7&92.0 & 100&98.7  \\
      \hline
              
          \end{tabular}
      \end{center}
  \vspace{-5mm}
  \caption{$\mathit{ASR}$ under different attacking conditions. $\widetilde{ASR}$ assesses the attack effectiveness of ``Trigger Optimization'' alone, while $\ddot{ASR}$ assesses the combined effectiveness of both ``Trigger Optimization'' and ``Concealment of Model Updates''.}
  \label{tbl:twoeffect}
  \vspace{-5mm}
\end{table}

\myparatight{Experiment results} Table~\ref{tbl:twoeffect} shows
results of $\widetilde{ASR}$ and $\ddot{ASR}$ over 10 different
defense methods. We used same
settings as in Table~\ref{tbl:defaultsetting} for testing
$\ddot{ASR}$, and kept the size of trigger training dataset
consistent when testing $\widetilde{ASR}$.

The results of $\widetilde{ASR}$ in Table~\ref{tbl:twoeffect} show
that different defense methods
resulted in very different $\widetilde{ASR}$ even for the same learning
task of a dataset. The reason for the variance of $\widetilde{ASR}$ is
the gap between ${W_g}^{(i)}$ and $\widetilde{W}_g^{(i+1)}$
were different when implementing different defense methods. According to the recent studies~\cite{CerP,zhang2024a3fl}, if the gap between consecutive rounds of global models in an FL system is smaller, Trigger Optimization will be more effective in its attack.

The results of $\ddot{ASR}$ in Table~\ref{tbl:twoeffect} show that
the presence of malicious clients' model updates consistently enhances $\mathit{ASR}$ compared to $\widetilde{ASR}$  across all defense methods on different datasets. We consider this enhancement as an evidence of the statement that the
attack effectiveness of \ourmodel{} comes from both Trigger
Optimization and Concealment of Model Updates, with the latter one playing a critical role in producing a high $\ddot{ASR}$.  

A general hypothesis made by the state-of-the-art defenses against
backdoor attacks in FL is that malicious clients' model updates have a distinct
divergence from benign clients' model updates. However, as indicated by the results in Table~\ref{tbl:twoeffect}, \ourmodel{} effectively conceals the model updates from malicious clients amidst those of benign clients, eluding detection and filtering by state-of-the-art defenses. Consequently, defenses formulated based on this broad hypothesis will inherently struggle to defend against \ourmodel{} attacks.

\subsubsection{Impact of Malicious Client Ratio (MCR)}


\begin{table*}[!h]

  \renewcommand{\arraystretch}{1.2}
  
\begin{center}

\scriptsize\addtolength{\tabcolsep}{-4.7pt}
\begin{tabular}{|c|ccc|ccc|ccc|ccc||ccc|ccc|ccc|ccc|}
  \hline
  &\multicolumn{12}{c||}{\footnotesize \textbf{Final $\mathit{ASR}$}} & \multicolumn{12}{c|}{\footnotesize \textbf{Average $\mathit{ASR}$}}  \\
  \hline
  \footnotesize \textbf{MCR}  &\multicolumn{3}{c|}{\textbf{0.05}} & \multicolumn{3}{c|}{\textbf{0.1}} & \multicolumn{3}{c|}{\textbf{0.2}}  &\multicolumn{3}{c||}{\textbf{0.3}}  &\multicolumn{3}{c|}{\textbf{0.05}} & \multicolumn{3}{c|}{\textbf{0.1}} & \multicolumn{3}{c|}{\textbf{0.2}}  &\multicolumn{3}{c|}{\textbf{0.3}}\\
  \hline
  & \scriptsize Ours & \scriptsize FT & \scriptsize  DFT & \scriptsize Ours & \scriptsize FT & \scriptsize DFT & \scriptsize Ours & \scriptsize FT & \scriptsize DFT & \scriptsize Ours & \scriptsize FT & \scriptsize DFT & \scriptsize Ours & \scriptsize FT & \scriptsize DFT & \scriptsize Ours & \scriptsize FT & \scriptsize DFT & \scriptsize Ours & \scriptsize FT & \scriptsize DFT & \scriptsize Ours & \scriptsize FT & \scriptsize DFT \\
  \hline
FedAvg & \textbf{100}	&\textbf{100}		&93			&\textbf{100}		&\textbf{100}		&\textbf{100}			&\textbf{100}		&\textbf{100}		&\textbf{100}			&\textbf{100}		&\textbf{100}		&\textbf{100}	 &\textbf{99} &88 &50 &\textbf{99} &96 &88 &\textbf{99} &\textbf{99} &92 &99 &\textbf{100} &97\\
\hline
Median &\textbf{100}		&81		&72			&\textbf{100}		&\textbf{100}		&97			&\textbf{100}		&\textbf{100}		&\textbf{100}			&\textbf{100}		&\textbf{100}		&\textbf{100}	     &\textbf{96} &47 &42 &\textbf{97} &79 &63 &\textbf{99} &97 &82 &\textbf{99} &98 &93\\
\hline
{\footnotesize Trimmed Mean} &\textbf{100}		&95		&38			&\textbf{100}		&\textbf{100}		&99			&\textbf{100}		&\textbf{100}		&\textbf{100}			&\textbf{100}		&\textbf{100}	&	\textbf{100}	      &\textbf{89} &59 &23 &\textbf{98} &82 &69 &\textbf{99} &94 &85 &\textbf{99} &\textbf{99} &92\\
\hline
RobustLR  &\textbf{100} &\textbf{100} &\textbf{100} &\textbf{100} &\textbf{100} &\textbf{100} &\textbf{100} &\textbf{100} &\textbf{100} &\textbf{100} &\textbf{100} &\textbf{100} &\textbf{99} &94 &87 &\textbf{99} &98 &94 &\textbf{99} &\textbf{99} &98 &\textbf{99} &\textbf{99} &\textbf{99}\\
\hline
RFA  &\textbf{100}  &\textbf{100}  &98  &\textbf{100}  &\textbf{100}  &\textbf{100}  &\textbf{100}  &\textbf{100} &\textbf{100}  &\textbf{100}  &\textbf{100}  &\textbf{100}       &\textbf{98} &81 &55 &\textbf{99} &95 &90 &\textbf{99}  &\textbf{99} &97 &\textbf{99} &\textbf{99} &98\\
\hline
FLAIR   &\textbf{62}  &15  &10  &\textbf{58}  &25  &9  &\textbf{67}  &27  &22  &\textbf{82}  &33  &40        &\textbf{51} &14 &10 &\textbf{64} &24 &9 &\textbf{68} &24 &16 &\textbf{84} &42 &30\\
\hline
FLCert &\textbf{99}		&93		&34			&\textbf{100}		&\textbf{100}		&95			&\textbf{100}		&\textbf{100}		&\textbf{100}			&\textbf{100}		&\textbf{100}		&\textbf{100}	       &\textbf{88} &60 &21 &\textbf{98} &87 &60 &\textbf{98} &94 &83 &\textbf{99} &\textbf{99} &91\\
\hline
FLAME &\textbf{60}		&19		&17			&\textbf{52}		&18		&51			&\textbf{50}		&16		&16			&\textbf{55}		&19		&16	       &\textbf{56} &18 &14&\textbf{66} &19 &34 &\textbf{53} &19 &16 &\textbf{70} &23 &43\\
\hline
FoolsGold &\textbf{100}		&\textbf{100}		&94			&\textbf{100}		&\textbf{100}		&\textbf{100}			&\textbf{100}		&\textbf{100}		&\textbf{100}		&\textbf{100}		&\textbf{100}	&\textbf{100}	       &\textbf{98} &88 &53 &\textbf{99} &97 &87 &\textbf{99} &\textbf{99} &95	&\textbf{99} &	\textbf{99} &98\\
\hline
Multi-Krum  &\textbf{100} &\textbf{100} &\textbf{100} &\textbf{100} &\textbf{100} &\textbf{100} &\textbf{100} &\textbf{100} &\textbf{100} &\textbf{100} &\textbf{100} &\textbf{100} &\textbf{99} &\textbf{99} &83 &99 &\textbf{100} &98 &98 &\textbf{100} &99 &99 &\textbf{100} &\textbf{100}\\
\hline

\end{tabular}
\end{center}
\vspace{-5mm}
\caption{The effects of malicious client ratio on the effectiveness of different attacks (CIFAR10).}
\label{tbl:MCR_ASR_CIFAR10}
\vspace{-2mm}
\end{table*}

\begin{table*}[!h]
  \renewcommand{\arraystretch}{1.2}
\begin{center}
  \scriptsize\addtolength{\tabcolsep}{-4.5pt}
  \begin{tabular}{|c|ccc|ccc|ccc|ccc||ccc|ccc|ccc|ccc|}
    \hline
    &\multicolumn{12}{c||}{\textbf{\footnotesize Final $\mathit{ASR}$}} & \multicolumn{12}{c|}{\footnotesize \textbf{Average $\mathit{ASR}$}}  \\
    \hline
    \footnotesize \textbf{Trigger Size}  &\multicolumn{3}{c|}{\textbf{9}} & \multicolumn{3}{c|}{\textbf{25}} & \multicolumn{3}{c|}{\textbf{49}}  &\multicolumn{3}{c||}{\textbf{100}} &\multicolumn{3}{c|}{\textbf{9}} & \multicolumn{3}{c|}{\textbf{25}} & \multicolumn{3}{c|}{\textbf{49}}  &\multicolumn{3}{c|}{\textbf{100}} \\
    \hline
    & \scriptsize Ours & \scriptsize FT & \scriptsize  DFT & \scriptsize Ours & \scriptsize FT & \scriptsize DFT & \scriptsize Ours & \scriptsize FT & \scriptsize DFT & \scriptsize Ours & \scriptsize FT & \scriptsize DFT & \scriptsize Ours & \scriptsize FT & \scriptsize DFT & \scriptsize Ours & \scriptsize FT & \scriptsize DFT & \scriptsize Ours & \scriptsize FT & \scriptsize DFT & \scriptsize Ours & \scriptsize FT & \scriptsize DFT \\
    \hline
FedAvg &\textbf{100}		&94		&49			&\textbf{100}		&\textbf{100}		&93			&\textbf{100}		&\textbf{100}		&91			&\textbf{100}		&\textbf{100}		&77	        &\textbf{95}&60 &28 &\textbf{99} &88 &50 &\textbf{99} &90 &59 &\textbf{99} &93 &52\\
\hline
Median &\textbf{97}		&23		&12			&\textbf{100}		&81		&72			&\textbf{100}		&95		&25			&\textbf{100}		&99		&46	        &\textbf{66} &21 &12 &\textbf{96} &47 &42 &\textbf{98} &66 &17 &\textbf{99} &82 &29\\
\hline
{\footnotesize Trimmed Mean} &\textbf{98}&51&14&\textbf{100}&95&38&\textbf{100}&99&43&\textbf{100}&\textbf{100}&74        &\textbf{71}&29&13&\textbf{89}&59&23&\textbf{99}&74 &27&\textbf{99} &79 &44\\
\hline
RobustLR  &\textbf{100} &\textbf{100} &\textbf{100} &\textbf{100} &\textbf{100} &\textbf{100} &\textbf{100} &\textbf{100} &98 &\textbf{100} &\textbf{100} &99 &\textbf{95} &91 &69 &\textbf{99} &94 &87 &\textbf{99} &94 &77 &\textbf{99} &95 &82 \\
\hline
RFA    &\textbf{100}  &\textbf{100}  &99  &\textbf{100}  &\textbf{100}  &98  &\textbf{100}  &\textbf{100}  &\textbf{100}  &\textbf{100}  &\textbf{100}  &98      &\textbf{93}&79&56 &\textbf{98}&81&55 &\textbf{99}&81&71&\textbf{99}&90&73 \\
\hline
FLAIR  &\textbf{27}  &14  &14  &\textbf{62}  &15  &10  &\textbf{89}  &22  &15  &\textbf{99}  &24  &14        &\textbf{24} &14 &13 &\textbf{51} &14 &10 &\textbf{84} &22 &15 &\textbf{98} &16 &13 \\
\hline
FLCert &\textbf{99}&38&14&\textbf{99}&93&34&\textbf{100}&88&51&\textbf{100}&\textbf{100}&49       &\textbf{78} &26&13 &\textbf{88} &60 &21 &\textbf{99} &59 &23 &\textbf{99}&78 &33\\
\hline
FLAME &\textbf{21}&18&12&\textbf{60}&19&17&\textbf{100}&12&11&\textbf{100}&33&31      &\textbf{35}&17&12 &\textbf{56}&18&14  &\textbf{84}&17&11 &\textbf{90}&31&24 \\
\hline
FoolsGold &\textbf{100}&\textbf{100}&43&\textbf{100}&\textbf{100}&94&\textbf{100}&\textbf{100}&98&\textbf{100}&\textbf{100}&81      &\textbf{93} &72&23 &\textbf{98} &88 &53 &\textbf{99} &94 &69 &\textbf{99} &94 &55\\
\hline
Multi-Krum  &\textbf{100} &\textbf{100} &15 &\textbf{100} &\textbf{100} &\textbf{100} &99 &\textbf{100} &\textbf{100} &\textbf{100} &\textbf{100} &\textbf{100} &\textbf{99} &\textbf{99} &11 &\textbf{99} &\textbf{99} &83 &\textbf{99} &\textbf{99} &95 &\textbf{99} &\textbf{99} &97\\
\hline

\end{tabular}
\end{center}
\vspace{-5mm}
\caption{The effects of trigger size on the effectiveness of different attacks (CIFAR10).}
\label{tbl:TS_ASR_CIFAR10}
\vspace{-5mm}
\end{table*}

In this section, we evaluated the impact of different Malicious Client
Ratios (MCR) on the attacking performance of \ourmodel{} attack. We assumed that the number of malicious clients
in the FL system should be kept small ($\leq 30\%$) for practical
reasons. We varied the MCR  across four different settings (0.05, 0.1, 0.2, and 0.3) while keeping other settings consistent with those in
Table~\ref{tbl:defaultsetting}. We experimented over 10 different
defenses on the learning tasks of the CIFAR10 datasets and
compare \ourmodel{}'s results with FT and DFT.
 
Tables~\ref{tbl:MCR_ASR_CIFAR10} presents the evaluation results of attack effectiveness. \ourmodel{} exhibited a dominant advantage
over FT and DFT when the MCR is small (0.05 and 0.1). However,
this advantage diminished with increasing MCR, indicating that
when a sufficient number of malicious clients present in FL, even
FT and DFT can achieve respectable $\mathit{ASR}$ against certain
defense strategies. In most cases, the $\mathit{ASR}$ for all attacks
continued to rise as the MCR increased, with the exception of
FLAME. Results obtained with FLAME indicate that the number of
malicious clients did not significantly impact its defense
effectiveness. 

Table~\ref{tbl:MCR_MA_CIFAR10} presents the Main-task Accuracy results for each experiment considered in this section. All MA results for different attacks remain similar to the baseline MA, indicating the correct implementation of each attack.

\subsubsection{Impact of Trigger Size}
\vspace{-2mm}

Trigger Size, determining how many pixels in an image we can alter, is an important parameter for \ourmodel{} attack. Larger trigger size generally results in a
better optimization performance.  However, it is essential to strike a
balance because an excessively large trigger size will make a trigger obscure important details of images, making the trigger easier to perceive by humans.  In this section, we assessed the impact of different
trigger sizes on the performance of different attacks. We explored trigger sizes across four
different settings (9, 25, 49, and 100) while maintaining other
settings in accordance with those outlined in
Table~\ref{tbl:defaultsetting}. 

Tables~\ref{tbl:TS_ASR_CIFAR10} shows that \ourmodel{} maintained a significant
advantage in $\mathit{ASR}$ over FT and DFT across various trigger
sizes, ranging from small to large. According to the results, we found that FT and DFT did not benefit
from larger trigger sizes in achieving higher
$\mathit{ASR}$ when encountering with robust aggregations that have advanced defense effectiveness, such as FLAIR and FLAME. A possible explanation on that is when malicious model updates were trained
on data poisoned with larger FT or DFT triggers, they exhibited greater
divergence from benign model updates, making them more susceptible to
detection and filtering by defense mechanisms. In
contrast, \ourmodel{} demonstrated a continuous improvement in
$\mathit{ASR}$ as the trigger size increased.

Table~\ref{tbl:TS_MA_CIFAR10} presents the Main-task Accuracy results for each experiment considered in this section. Results in it indicate all backdoor attacks achieved their stealthiness goals during attacking.

\section{Conclusion and Future Work}

In this work, we proposed \ourmodel{}, a novel backdoor attack method in federated learning (FL). \ourmodel{} dynamically adjusts the backdoor objective to conceal malicious clients' model updates among benign ones, enabling global models to aggregate them even when protected by state-of-the-art defenses. \ourmodel{} attack is easy to implement, relying solely on data poisoning, yet it poses a significant threat to existing defense methods. 

Future work based on this paper includes extending the research to other learning tasks beyond image classification, such as text generation. Additionally, the time and computational costs of implementing our attack were not discussed, as we assumed attackers could use more powerful resources; thus, optimizing these aspects and developing timing-based defenses will be explored later. Lastly, designing defenses against backdoor attacks like \ourmodel{} will need to account for scenarios where a malicious client's model is indistinguishable from its non-attacked model. It is also crucial to ensure that defenses adhere to FL's privacy-preserving principles, in line with the primitive version that attracted users to FL.


\bibliographystyle{plain}
\bibliography{ref.bib} 

\begin{thebibliography}{10}

\bibitem{howtobackdoor}
Eugene Bagdasaryan, Andreas Veit, Yiqing Hua, Deborah Estrin, and Vitaly Shmatikov.
\newblock How to backdoor federated learning.
\newblock In {\em International Conference on Artificial Intelligence and Statistics}, pages 2938--2948. PMLR, 2020.

\bibitem{alittleisenough}
Gilad Baruch, Moran Baruch, and Yoav Goldberg.
\newblock A little is enough: Circumventing defenses for distributed learning.
\newblock {\em Advances in Neural Information Processing Systems}, 32, 2019.

\bibitem{krum}
Peva Blanchard, El~Mahdi El~Mhamdi, Rachid Guerraoui, and Julien Stainer.
\newblock Machine learning with adversaries: Byzantine tolerant gradient descent.
\newblock {\em Advances in neural information processing systems}, 30, 2017.

\bibitem{caldas2018leaf}
Sebastian Caldas, Sai Meher~Karthik Duddu, Peter Wu, Tian Li, Jakub Kone{\v{c}}n{\`y}, H~Brendan McMahan, Virginia Smith, and Ameet Talwalkar.
\newblock Leaf: A benchmark for federated settings.
\newblock {\em arXiv preprint arXiv:1812.01097}, 2018.

\bibitem{cao2020fltrust}
Xiaoyu Cao, Minghong Fang, Jia Liu, and Neil~Zhenqiang Gong.
\newblock Fltrust: Byzantine-robust federated learning via trust bootstrapping.
\newblock {\em arXiv preprint arXiv:2012.13995}, 2020.

\bibitem{cao2022flcert}
Xiaoyu Cao, Zaixi Zhang, Jinyuan Jia, and Neil~Zhenqiang Gong.
\newblock Flcert: Provably secure federated learning against poisoning attacks.
\newblock {\em IEEE Transactions on Information Forensics and Security}, 17:3691--3705, 2022.

\bibitem{chen2017targeted}
Xinyun Chen, Chang Liu, Bo~Li, Kimberly Lu, and Dawn Song.
\newblock Targeted backdoor attacks on deep learning systems using data poisoning.
\newblock {\em arXiv preprint arXiv:1712.05526}, 2017.

\bibitem{fang2020local}
Minghong Fang, Xiaoyu Cao, Jinyuan Jia, and Neil Gong.
\newblock Local model poisoning attacks to $\{$Byzantine-Robust$\}$ federated learning.
\newblock In {\em 29th USENIX security symposium (USENIX Security 20)}, pages 1605--1622, 2020.

\bibitem{fang2024byzantine}
Minghong Fang, Zifan Zhang, Prashant Khanduri, Songtao Lu, Yuchen Liu, Neil Gong, et~al.
\newblock Byzantine-robust decentralized federated learning.
\newblock {\em arXiv preprint arXiv:2406.10416}, 2024.

\bibitem{fang2023vulnerability}
Pei Fang and Jinghui Chen.
\newblock On the vulnerability of backdoor defenses for federated learning.
\newblock In {\em Proceedings of the AAAI Conference on Artificial Intelligence}, volume~37, pages 11800--11808, 2023.

\bibitem{freqfed}
Hossein Fereidooni, Alessandro Pegoraro, Phillip Rieger, Alexandra Dmitrienko, and Ahmad-Reza Sadeghi.
\newblock Freqfed: A frequency analysis-based approach for mitigating poisoning attacks in federated learning.
\newblock {\em arXiv preprint arXiv:2312.04432}, 2023.

\bibitem{foolsgold}
Clement Fung, Chris J.~M. Yoon, and Ivan Beschastnikh.
\newblock The limitations of federated learning in sybil settings.
\newblock In {\em 23rd International Symposium on Research in Attacks, Intrusions and Defenses (RAID 2020)}, pages 301--316, San Sebastian, October 2020. USENIX Association.

\bibitem{fung2018mitigating}
Clement Fung, Chris~JM Yoon, and Ivan Beschastnikh.
\newblock Mitigating sybils in federated learning poisoning.
\newblock {\em arXiv preprint arXiv:1808.04866}, 2018.

\bibitem{gong2022coordinated}
Xueluan Gong, Yanjiao Chen, Huayang Huang, Yuqing Liao, Shuai Wang, and Qian Wang.
\newblock Coordinated backdoor attacks against federated learning with model-dependent triggers.
\newblock {\em IEEE network}, 36(1):84--90, 2022.

\bibitem{gu2019badnets}
Tianyu Gu, Kang Liu, Brendan Dolan-Gavitt, and Siddharth Garg.
\newblock Badnets: Evaluating backdooring attacks on deep neural networks.
\newblock {\em IEEE Access}, 7:47230--47244, 2019.

\bibitem{he2016deep}
Kaiming He, Xiangyu Zhang, Shaoqing Ren, and Jian Sun.
\newblock Deep residual learning for image recognition.
\newblock In {\em Proceedings of the IEEE conference on computer vision and pattern recognition}, pages 770--778, 2016.

\bibitem{kabir2023flshield}
Ehsanul Kabir, Zeyu Song, Md~Rafi~Ur Rashid, and Shagufta Mehnaz.
\newblock Flshield: A validation based federated learning framework to defend against poisoning attacks.
\newblock {\em arXiv preprint arXiv:2308.05832}, 2023.

\bibitem{kolouri2020universal}
Soheil Kolouri, Aniruddha Saha, Hamed Pirsiavash, and Heiko Hoffmann.
\newblock Universal litmus patterns: Revealing backdoor attacks in cnns.
\newblock In {\em Proceedings of the IEEE/CVF Conference on Computer Vision and Pattern Recognition}, pages 301--310, 2020.

\bibitem{baybfed}
K.~Kumari, P.~Rieger, H.~Fereidooni, M.~Jadliwala, and A.~Sadeghi.
\newblock Baybfed: Bayesian backdoor defense for federated learning.
\newblock In {\em 2023 2023 IEEE Symposium on Security and Privacy (SP) (SP)}, 2023.

\bibitem{lin2020composite}
Junyu Lin, Lei Xu, Yingqi Liu, and Xiangyu Zhang.
\newblock Composite backdoor attack for deep neural network by mixing existing benign features.
\newblock In {\em Proceedings of the 2020 ACM SIGSAC Conference on Computer and Communications Security}, pages 113--131, 2020.

\bibitem{liu2018trojaning}
Yingqi Liu, Shiqing Ma, Yousra Aafer, Wen-Chuan Lee, Juan Zhai, Weihang Wang, and Xiangyu Zhang.
\newblock Trojaning attack on neural networks.
\newblock In {\em 25th Annual Network And Distributed System Security Symposium (NDSS 2018)}. Internet Soc, 2018.

\bibitem{CerP}
Xiaoting Lyu, Yufei Han, Wei Wang, Jingkai Liu, Bin Wang, Jiqiang Liu, and Xiangliang Zhang.
\newblock Poisoning with cerberus: Stealthy and colluded backdoor attack against federated learning.
\newblock {\em Proceedings of the AAAI Conference on Artificial Intelligence}, 37:9020--9028, Jun. 2023.

\bibitem{mcmahan2017fl}
Brendan McMahan, Eider Moore, Daniel Ramage, Seth Hampson, and Blaise~Aguera y~Arcas.
\newblock Communication-efficient learning of deep networks from decentralized data.
\newblock In {\em Artificial intelligence and statistics}, pages 1273--1282. PMLR, 2017.

\bibitem{topological}
Xiaoxing Mo, Yechao Zhang, Leo~Yu Zhang, Wei Luo, Nan Sun, Shengshan Hu, Shang Gao, and Yang Xiang.
\newblock Robust backdoor detection for deep learning via topological evolution dynamics.
\newblock In {\em 2024 IEEE Symposium on Security and Privacy (SP)}, pages 171--171. IEEE Computer Society, 2024.

\bibitem{everyvotecounts}
Hamid Mozaffari, Virat Shejwalkar, and Amir Houmansadr.
\newblock Every vote counts:$\{$Ranking-Based$\}$ training of federated learning to resist poisoning attacks.
\newblock In {\em 32nd USENIX Security Symposium (USENIX Security 23)}, pages 1721--1738, 2023.

\bibitem{nguyen2022flame}
Thien~Duc Nguyen, Phillip Rieger, Roberta De~Viti, Huili Chen, Bj{\"o}rn~B Brandenburg, Hossein Yalame, Helen M{\"o}llering, Hossein Fereidooni, Samuel Marchal, Markus Miettinen, et~al.
\newblock $\{$FLAME$\}$: Taming backdoors in federated learning.
\newblock In {\em 31st USENIX Security Symposium (USENIX Security 22)}, pages 1415--1432, 2022.

\bibitem{robustlr}
Mustafa~Safa Ozdayi, Murat Kantarcioglu, and Yulia~R Gel.
\newblock Defending against backdoors in federated learning with robust learning rate.
\newblock In {\em Proceedings of the AAAI Conference on Artificial Intelligence}, 2021.

\bibitem{RFA}
Krishna Pillutla, Sham~M. Kakade, and Zaid Harchaoui.
\newblock Robust aggregation for federated learning.
\newblock {\em IEEE Transactions on Signal Processing}, 70:1142--1154, 2022.

\bibitem{rieger2022crowdguard}
Phillip Rieger, Torsten Krau{\ss}, Markus Miettinen, Alexandra Dmitrienko, and Ahmad-Reza Sadeghi.
\newblock Crowdguard: Federated backdoor detection in federated learning.
\newblock {\em arXiv preprint arXiv:2210.07714}, 2022.

\bibitem{sandeepa2024sherpa}
Chamara Sandeepa, Bartlomiej Siniarski, Shen Wang, and Madhusanka Liyanage.
\newblock Sherpa: Explainable robust algorithms for privacy-preserved federated learning in future networks to defend against data poisoning attacks.
\newblock In {\em 2024 IEEE Symposium on Security and Privacy (SP)}, pages 204--204. IEEE Computer Society, 2024.

\bibitem{schneider:2022:sok}
Moritz Schneider, Ramya~Jayaram Masti, Shweta Shinde, Srdjan Capkun, and Ronald Perez.
\newblock {SoK}: Hardware-supported trusted execution environments.
\newblock {\em arXiv preprint arXiv:2205.12742}, 2022.

\bibitem{shafahi2018poison}
Ali Shafahi, W~Ronny Huang, Mahyar Najibi, Octavian Suciu, Christoph Studer, Tudor Dumitras, and Tom Goldstein.
\newblock Poison frogs! targeted clean-label poisoning attacks on neural networks.
\newblock {\em Advances in neural information processing systems}, 31, 2018.

\bibitem{flair}
Atul Sharma, Wei Chen, Joshua Zhao, Qiang Qiu, Saurabh Bagchi, and Somali Chaterji.
\newblock Flair: Defense against model poisoning attack in federated learning.
\newblock In {\em ASIA CCS '23}. Association for Computing Machinery, 2023.

\bibitem{simonyan2014very}
Karen Simonyan and Andrew Zisserman.
\newblock Very deep convolutional networks for large-scale image recognition.
\newblock {\em arXiv preprint arXiv:1409.1556}, 2014.

\bibitem{sun2019can}
Ziteng Sun, Peter Kairouz, Ananda~Theertha Suresh, and H~Brendan McMahan.
\newblock Can you really backdoor federated learning?
\newblock {\em arXiv preprint arXiv:1911.07963}, 2019.

\bibitem{wang2019neural}
Bolun Wang, Yuanshun Yao, Shawn Shan, Huiying Li, Bimal Viswanath, Haitao Zheng, and Ben~Y Zhao.
\newblock Neural cleanse: Identifying and mitigating backdoor attacks in neural networks.
\newblock In {\em 2019 IEEE Symposium on Security and Privacy (SP)}, pages 707--723. IEEE, 2019.

\bibitem{yes}
Hongyi Wang, Kartik Sreenivasan, Shashank Rajput, Harit Vishwakarma, Saurabh Agarwal, Jy-yong Sohn, Kangwook Lee, and Dimitris Papailiopoulos.
\newblock Attack of the tails: Yes, you really can backdoor federated learning.
\newblock In {\em Proceedings of the 34th International Conference on Neural Information Processing Systems}, NIPS'20, 2020.

\bibitem{wei2022vertical}
Kang Wei, Jun Li, Chuan Ma, Ming Ding, Sha Wei, Fan Wu, Guihai Chen, and Thilina Ranbaduge.
\newblock Vertical federated learning: Challenges, methodologies and experiments.
\newblock {\em arXiv preprint arXiv:2202.04309}, 2022.

\bibitem{Xie2020DBA:}
Chulin Xie, Keli Huang, Pin-Yu Chen, and Bo~Li.
\newblock Dba: Distributed backdoor attacks against federated learning.
\newblock In {\em International Conference on Learning Representations}, 2020.

\bibitem{xie2022efficient}
Kan Xie, Zhe Zhang, Bo~Li, Jiawen Kang, Dusit Niyato, Shengli Xie, and Yi~Wu.
\newblock Efficient federated learning with spike neural networks for traffic sign recognition.
\newblock {\em IEEE Transactions on Vehicular Technology}, 71(9):9980--9992, 2022.

\bibitem{yao2019latent}
Yuanshun Yao, Huiying Li, Haitao Zheng, and Ben~Y Zhao.
\newblock Latent backdoor attacks on deep neural networks.
\newblock In {\em Proceedings of the 2019 ACM SIGSAC conference on computer and communications security}, pages 2041--2055, 2019.

\bibitem{trim_median}
Dong Yin, Yudong Chen, Ramchandran Kannan, and Peter Bartlett.
\newblock Byzantine-robust distributed learning: Towards optimal statistical rates.
\newblock In {\em International Conference on Machine Learning}, pages 5650--5659. PMLR, 2018.

\bibitem{yueqifedredefense}
XIE Yueqi, Minghong Fang, and Neil~Zhenqiang Gong.
\newblock Fedredefense: Defending against model poisoning attacks for federated learning using model update reconstruction error.
\newblock In {\em Forty-first International Conference on Machine Learning}, 2024.

\bibitem{zhang2024a3fl}
Hangfan Zhang, Jinyuan Jia, Jinghui Chen, Lu~Lin, and Dinghao Wu.
\newblock A3fl: Adversarially adaptive backdoor attacks to federated learning.
\newblock {\em Advances in Neural Information Processing Systems}, 36, 2024.

\bibitem{zhang2022flip}
Kaiyuan Zhang, Guanhong Tao, Qiuling Xu, Siyuan Cheng, Shengwei An, Yingqi Liu, Shiwei Feng, Guangyu Shen, Pin-Yu Chen, Shiqing Ma, et~al.
\newblock Flip: A provable defense framework for backdoor mitigation in federated learning.
\newblock {\em arXiv preprint arXiv:2210.12873}, 2022.

\bibitem{zhang2022neurotoxin}
Zhengming Zhang, Ashwinee Panda, Linyue Song, Yaoqing Yang, Michael Mahoney, Prateek Mittal, Ramchandran Kannan, and Joseph Gonzalez.
\newblock Neurotoxin: Durable backdoors in federated learning.
\newblock In {\em International Conference on Machine Learning}, pages 26429--26446. PMLR, 2022.

\end{thebibliography}

\clearpage
\appendix
\appendixpage
\section{Additional Related Works}

\subsection{Clean-label attacks}
Clean-label attacks~\cite{shafahi2018poison} involve manipulating input data with subtle perturbations while keeping labels unchanged. Although this assumption aligns with scenarios like Vertical Federated Learning ~\cite{wei2022vertical} (VFL), where participants possess vertically partitioned data with labels owned by only one participant, our study does not consider VFL as our attack scenario. Furthermore, we focus on examining the effects of different backdoor triggers on hiding malicious model updates rather than their imperceptible characteristics. Therefore, discussions of clean-label attacks are beyond the scope of our work.

\subsection{Defenses with different privacy-preserving properties}\label{apx:related_work}

Recent defense works have introduced several unconventional FL pipelines aimed at enhancing the security of FL against various types of attacks. These novel architectures provide different levels of privacy protection and often require additional techniques (e.g., Secured Multi-party Computation) to ensure privacy for FL clients. In light of these privacy considerations, we have chosen to focus our analysis on the conventional FL structure that was originally proposed in the concept of Federated Learning~\cite{mcmahan2017fl}. Although defenses built on newly proposed FL structures fall outside the scope of our main comparison, we offer a discussion of these related works in this section.

\myparatight{Clients' private data were shared to the server}
Some approaches allow the server to have access to a small portion of main-task data shared by clients. To mitigate backdoor attacks, server-side defense strategies use this data to either independently train a model and use its updates as a reference for each round of aggregation (e.g., FLTrust~\cite{cao2020fltrust}), or to validate clients' model updates and eliminate those with abnormal outputs (e.g., SSDT~\cite{topological}, SHERPA~\cite{sandeepa2024sherpa}). However, both of these methods still rely on analyzing clients' model updates, making them vulnerable to backdoor attacks with dynamic objectives that conceal malicious updates. FedREdefense~\cite{yueqifedredefense} detects and filters out artificial model updates by reconstructing distilled data shared by clients, but this approach is not effective against backdoor attacks where malicious clients genuinely train their models on poisoned local data rather than fabricating artificial updates.

\myparatight{Clients' model updates were shared to each other}
Some approaches propose allowing clients to share their model updates with one another, rather than just with the server. CrowdGuard~\cite{rieger2022crowdguard} and FLShield~\cite{kabir2023flshield} suggest that a subset of clients validate other clients' model updates using their own data, assuming that malicious clients' updates would produce abnormal outputs on benign data. However, this hypothesis fails when malicious clients' updates are indistinguishable from non-backdoored updates, a state that can be achieved through backdoor attack with optimized triggers. Fang et al.~\cite{fang2024byzantine} proposed a decentralized FL framework without a central server, where clients exchange model updates and apply Byzantine-robust aggregation using their own updates as a reference. Like other defenses that rely on analyzing clients' model updates, this approach is also vulnerable to backdoor attacks with optimized triggers.

\section{Proofs} 

\subsection{Proof of Proposition~\ref{thrm: cossim}}

\begin{proof} \label{proof: cossim}
  The gradient of benign loss \eqnref{eq:bnloss} with respect to $\beta$ is
    \begin{equation*}
        g_{bn} = \frac{\partial L(x,\hat{y})}{\partial \beta} = 2x^T(x\beta-\hat{y}).
    \end{equation*}\label{eq:g_bd}
    The gradient of backdoor loss \eqnref{eq:bdloss} with respect to $\beta$ is
    \begin{equation*}
        g_{bd} = \frac{\partial  L(x_t,y_t)}{\partial \beta} =  2x_t^T(x_t\beta - y_t)
    \end{equation*}
    Gradients $G_{bn}$ and $G_{bd}$ defined by \eqnref{eq:G_bn} and \eqnref{eq:G_bd} can be written as
    \begin{align*}
        G_{bn} &= g_{bn}. \\
        G_{bd} &= (1-\alpha)g_{bn} + \alpha g_{bd}.
    \end{align*}

    The cosine similarity between $G_{bn}$ and $G_{bd}$ is
    \begin{align*}
        CosSim(G_{bn},G_{bd}) & = \frac{g_{bn}\cdot((1-\alpha)g_{bn} + \alpha g_{bd})}{\mid g_{bn}\mid\cdot\mid (1-\alpha)g_{bn} + \alpha g_{bd}\mid} \\
        & = \frac{g_{bn}\cdot(g_{bn} + \frac{\alpha}{1-\alpha} g_{bd})}{\mid g_{bn}\mid\cdot\mid g_{bn} + \frac{\alpha}{1-\alpha}g_{bd}\mid}
    \end{align*}
    
    One sufficiency to maximize $CosSim(G_{bn},G_{bd})$ is to minimize the distance between $g_{bn}$ and $g_{bn} + \frac{\alpha}{1-\alpha}g_{bd}$, which is 
    \begin{align*}
        \Delta d & = \mid g_{bn} - (g_{bn} + \frac{\alpha}{1-\alpha}g_{bd})\mid \\
        & = \frac{\alpha}{1-\alpha}\mid g_{bd}\mid.
    \end{align*}
    
    Since $\alpha$ is a constant, minimizing $\Delta d$ is equivalent to minimizing $\mid g_{bd} \mid$, which is bounded by
    \begin{equation*}
        0 \leq \mid g_{bd} \mid = \mid 2x_t^T(x_t\beta - y_t) \mid \leq 2 \mid e^T \mid \cdot \mid x_t\beta - y_t \mid ,
    \end{equation*}
    where $e^T \in \mathbb{R}^{1\times n}$ consists of the largest edge of the domain of $x_t$, e.g. $\mathbf{1}^T$ if considering normalization.   
    
    Thus, the optimization objective is to decrease $\mid g_{bd} \mid$ by minimizing its upper bound.
    \begin{align*}
        \min \mid x_t\beta - y_t \mid,
    \end{align*}
    which can be achieved by 
    \begin{align*} 
        \min_{V_t,E_t} & \parallel (x (I_n - E_t) + V_t E_t)\beta - y_t \parallel^2_2. 
    \end{align*}
\end{proof}

\subsection{Proposition \ref{thrm:2}}
\begin{proposition}\label{thrm:2}
  For any fixed trigger $\tau_f(V_t, E_t, y_t)$ with specified trigger
  value $V_t$, trigger location $E_t$, and predicted value $y_t$, there
  exists an optimal backdoor trigger $\hat{\tau}(\hat{V_t}, E_t, y_t)$
  that has the same $E_t$ and $y_t$ but optimizes its $V_t$
  with respect to a model $\beta$, which can result in a smaller
  or equal backdoor loss on model $\beta$ compared to $\tau_f$.
  \end{proposition}

\begin{proof} \label{proof:2}
  With a specified location $E_t$ and predicted value $y_t$, the optimization objective for minimizing backdoor loss is
  \begin{equation*}
      f = \min_{V_t} \parallel (x (I_n - E_t) + V_t E_t)\beta - y_t \parallel^2_2  
  \end{equation*}
  Since $V_t\in\mathcal{D}^{1\times n}$ where $\mathcal{D}$ is a convex domain and $\frac{\partial^2 f}{\partial V_t^2} \succeq 0$ for any $V_t \in \mathcal{D}^{1\times n}$, $f:\mathcal{D}^{1\times n}\rightarrow \mathbb{R}$ is a convex function. Thus, there exists an optimal value $\hat{V_t}$ for the objective function $f$ in the domain $\mathcal{D}^{1\times n}$.
\end{proof}

\subsection{Proof of Proposition \ref{thrm:3}}
\begin{proof} \label{proof:3}
  Assume the value of data $x$ before embedding a trigger is $[x_1, x_2, ..., x_n]$. If an entry location in $x$ is able to reduce the backdoor loss of $\beta$ by optimizing its entry value more effectively than any individual entry location within $E_t$,  we incorporate this location into $\hat{E_t}$. After constructing a $\hat{E_t}$, we optimize value of entries within $\hat{E_t}$ to obtain the optimized trigger $\hat{\tau}$. We are going to prove that constructing the trigger location $\hat{E_t}$ in this way results in the optimized trigger $\hat{\tau}$ always outperforming the fixed trigger $\tau_f$ in terms of backdoor loss.

  We use $k$ to represent the number of trigger entries that have been embedded into $x$. Assume the trigger value $V_t$ is composed of $[v_1, v_2,..., v_n]$.
  
  When $k=0$, the backdoor loss is 
  \[L(x,y_t)_{k=0} = \parallel x\beta - y_t \parallel^2_2.\]
  
  When $k=1$, we calculate a location of interest $i$ by taking the largest absolute gradient of the $L(x,y_t)_{k=0}$ with respect to all entry locations in $x$, 
  \[i = arg\max \mid \frac{\partial L(x,y_t)_{k=0}}{\partial x_i} \mid.\]
  
  If the entry location $i$ is inside of $E_t$, according to Proposition~\ref{thrm:2}, there exists an optimal entry value $\hat{v_i}$ resulting in a smaller or equal backdoor loss compared to ${v_i}$. In this case, we save $i$ as one of entry location in $\hat{E_t}$.

  If the entry location $i$ is outside of $E_t$, we have the following observation:

 For any entry location $j$ inside of $E_t$, we already know 
  \[\mid \frac{\partial L(x,y_t)_{k=0}}{\partial x_i} \mid \geq \mid \frac{\partial L(x,y_t)_{k=0}}{\partial x_j} \mid.\]

  We use Gradient Descent optimization algorithm to decrease loss by updating the entry value of the selected location with a constant step size $\Delta v$. When the selected location is $i$, the updated loss $L(x^{{i}},y_t)_{k=1}$ will be
  \[L(x^{\{i\}},y_t)_{k=1} = L(x,y_t)_{k=0} - \frac{\partial L(x,y_t)_{k=0}}{\partial x_i} \Delta v,\]
  and when the selected location is $j$, it is
  \[L(x^{\{j\}},y_t)_{k=1} = L(x,y_t)_{k=0} - \frac{\partial L(x,y_t)_{k=0}}{\partial x_j} \Delta v.\]
  
  It can be found that 
  \[L(x^{\{i\}},y_t)_{k=1} \leq L(x^{\{j\}},y_t)_{k=1}.\]

  Therefore,  $i$ is a better entry location in reducing backdoor loss compared to $j$ when we constrain the updating step size $\Delta v$ being static. After repeating the optimization step iteratively, if we finally find the optimal entry value $\hat{v_i}$ resulting in a smaller backdoor loss compared to $\hat{v_j}$, then it must also outperform the fixed value $v_j$ in $V_t$ according to Proposition~\ref{thrm:2}. If so, we save $i$ as one of entry location in $\hat{E_t}$.  Otherwise, we save $j$ as one of location in $\hat{E_t}$.

  By recursively operating the procedures across $k=2,3,...$, we will finally construct a $\hat{E_t}$ in which every entry location is proved to contribute a better attack performance than entry locations defined in $E_t$. 

\end{proof}

\section{Descriptions of Defenses} \label{defensedescription}

We implement our attack on FL systems integrated with 9 different defense strategies and provide a brief introduction for each of them:

\textbf{FedAvg~\cite{mcmahan2017fl}}, a basic aggregation rule in FL, computes global model updates by averaging all clients' model updates. Despite its effectiveness on the main task, it is not robust enough to defend against backdoor attacks in the FL system.

\textbf{Median~\cite{trim_median}}, a simple but robust alternative to FedAvg, constructs the global model updates by taking the median of the values of model updates across all clients

\textbf{Trimmed Mean~\cite{trim_median}}, in our implementation, excludes the $40\%$ largest and $40\%$ smallest values of each parameter among all clients' model updates and takes the mean of the remaining $20\%$ as the global model updates.

\textbf{Multi-Krum~\cite{krum}} identifys an honest client whose model updates have the smallest Euclidean distance to all other clients' model updates and takes this honest client's model updates as the global model updates. Despite its robustness to prevent the FL system from being compromised by a minor number of adversaries, Multi-Krum is not able to ensure the convergence performance of the FL system on its main task when the data distribution of clients is highly non-IID.

\textbf{RobustLR~\cite{robustlr}} adjusts the aggregation server's learning rate, per dimension and per round, based on the sign information of
clients' updates.

\textbf{RFA~\cite{RFA}} computes a geometric median of clients' model updates and assigns weight factors to clients depending on their distance from the geometric median. Subsequently, it computes the weighted average of all clients' model updates to generate the global model updates.

\textbf{FLAIR~\cite{flair}} assigns different weight factors to clients according to the similarity of the coefficient signs  between client model updates and global model updates of the previous round,  and then takes the weighted average of all clients' model updates to form the global model updates. FLAIR requires the knowledge of exact number of malicious clients existing in the FL system.

\textbf{FLCert~\cite{cao2022flcert}} randomly clusters clients, calculates the median of model updates within each cluster, incorporates them into the previous round's global model, and derives the majority inference outcome from these cluster-updated global models as the final inference result for the entire FL system. In our implementation, we cluster clients into 5 groups, use FLCert inference outcome for testing the Attack Success Rate, and employ Median as the aggregation rule for updating the global model in each round.

\textbf{FLAME~\cite{nguyen2022flame}} first clusters clients' model updates according to their cosine similarity to each other, and then aggregates the clipped model updates within the largest cluster as the global model updates. 

\textbf{FoolsGold~\cite{fung2018mitigating}} reduces aggregation weights of a set of clients whose model updates constantly exhibit high cosine similarity to each other.

\begin{figure}[t]
    \begin{minipage}[c]{0.45\textwidth}
        \caption{FT trigger on Tiny ImageNet data. Training Data \ref{sfig:a} and \ref{sfig:e} are from different malicious clients. Test Data \ref{sfig:b} is used to test ASR.}
    \centering
    \subfloat[\footnotesize Training Data]{\label{sfig:a}\includegraphics[width=.3\textwidth]{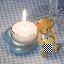}}\hfill
    \subfloat[\footnotesize Training Data]{\label{sfig:e}\includegraphics[width=.3\textwidth]{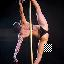}}\hfill
    \subfloat[\footnotesize Test Data]{\label{sfig:b}\includegraphics[width=.3\textwidth]{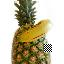}}
    \label{fig:poisoned_ft_ti}
    \end{minipage}

    \begin{minipage}[c]{0.45\textwidth}
        \caption{DFT trigger on Tiny ImageNet data. Training Data \ref{sfig:f} and \ref{sfig:c} are from different malicious clients. Test Data \ref{sfig:g} is used to test ASR.}
    \centering
    \subfloat[\footnotesize Training Data]{\label{sfig:f}\includegraphics[width=.3\textwidth]{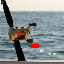}}\hfill
    \subfloat[\footnotesize Training Data]{\label{sfig:c}\includegraphics[width=.3\textwidth]{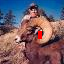}}\hfill
    \subfloat[\footnotesize Test Data]{\label{sfig:g}\includegraphics[width=.3\textwidth]{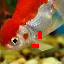}}
    \label{fig:poisoned_dft_ti}
    \end{minipage}

    \begin{minipage}[c]{0.45\textwidth}
        \caption{\ourmodel{} trigger on Tiny ImageNet data. Training Data \ref{sfig:d} and \ref{sfig:i} are from different malicious clients. Test Data \ref{sfig:h} is used to test ASR.}
        \centering
    \subfloat[\footnotesize Training Data]{\label{sfig:d}\includegraphics[width=.3\textwidth]{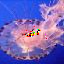}}\hfill
    \subfloat[\footnotesize Training Data]{\label{sfig:i}\includegraphics[width=.3\textwidth]{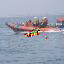}}\hfill
    \subfloat[\footnotesize Test Data]{\label{sfig:h}\includegraphics[width=.3\textwidth]{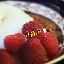}}
    \label{fig:poisoned_opt_ti}
    \end{minipage}

\end{figure}

\section{Visualization of Triggers}

\subsection{Different types of trigger on images}

We displayed different types of triggers on images from the Tiny ImageNet dataset in Figures \ref{fig:poisoned_opt_ti}, \ref{fig:poisoned_ft_ti}, and \ref{fig:poisoned_dft_ti}. The pattern of the FT trigger remains consistent across all datasets. The DFT triggers shown in Figure \ref{fig:poisoned_dft_ti} are the same as those used for images from the CIFAR10 dataset, while for the Fashion MNIST and FEMNIST datasets, DFT triggers appear in black.

\subsection{\ourmodel{} triggers on images from different datasets.}

\begin{figure*}[!t]
    \begin{minipage}[c]{0.80\textwidth}
    \begin{minipage}[c]{0.22\textwidth}
    \centering
    \subfloat{\includegraphics[width=.8\textwidth]{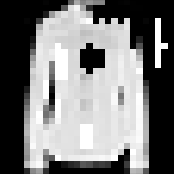}}\\
    \subfloat{\includegraphics[width=.8\textwidth]{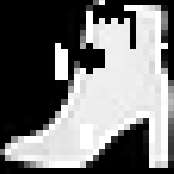}}\\
     Fashion MNIST
    \end{minipage}
    \hfill
    \begin{minipage}[c]{0.22\textwidth}
        \centering
    \subfloat{\includegraphics[width=.8\textwidth]{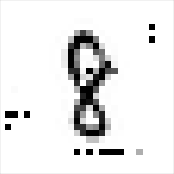}}\\
    \subfloat{\includegraphics[width=.8\textwidth]{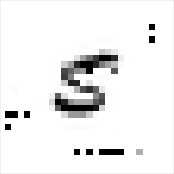}}\\
    FEMNIST
    \end{minipage}
    \hfill
    \begin{minipage}[c]{0.22\textwidth}
        \centering
    \subfloat{\includegraphics[width=.8\textwidth]{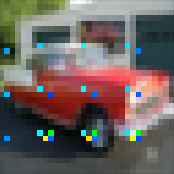}}\\
    \subfloat{\includegraphics[width=.8\textwidth]{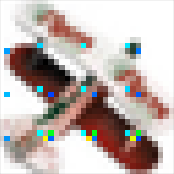}}\\
    CIFAR10
    \end{minipage}
    \hfill
    \begin{minipage}[c]{0.22\textwidth}
        \centering
    \subfloat{\includegraphics[width=.8\textwidth]{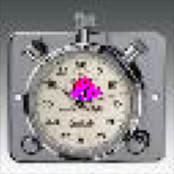}}\\
    \subfloat{\includegraphics[width=.8\textwidth]{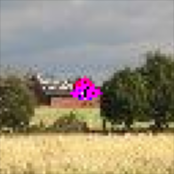}}\\
    Tiny ImageNet
    \end{minipage}
    \caption{\ourmodel{} triggers on images from different datasets.}
    \label{fig:poisoned_opt}
    \end{minipage}
    \hfill
    \begin{minipage}[c]{0.19\textwidth}
    \centering
    \subfloat{\includegraphics[width=.8\textwidth]{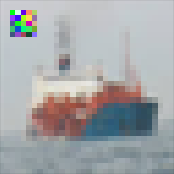}}\\
    \subfloat{\includegraphics[width=.8\textwidth]{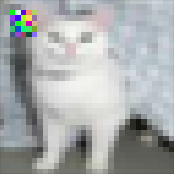}}\hfill
    \caption{A3FL trigger on images from CIFAR10.}
    \label{fig:poisoned_A3FL}
    \end{minipage}
\end{figure*}

We displayed \ourmodel{} triggers generated for images from different dataset in Figure~\ref{fig:poisoned_opt}. Our triggers are in a small size that could not obscure important details of any images.

\subsection{A3FL trigger on images of CIFAR10}

In Figure~\ref{fig:poisoned_A3FL}, we showed triggers generated by A3FL's methods on images from CIFAR10.

\subsection{Trigger evolution during training}

\begin{figure*}[htbp]
    \centering
    \subfloat[Round 10]{\includegraphics[width=.18\textwidth]{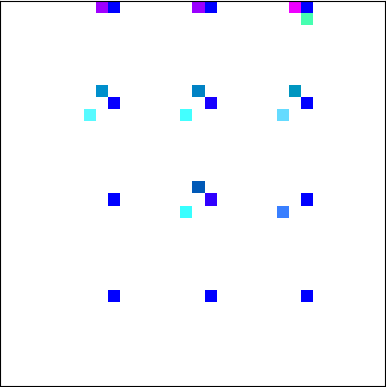}}\hfill
    \subfloat[Round 20]{\includegraphics[width=.18\textwidth]{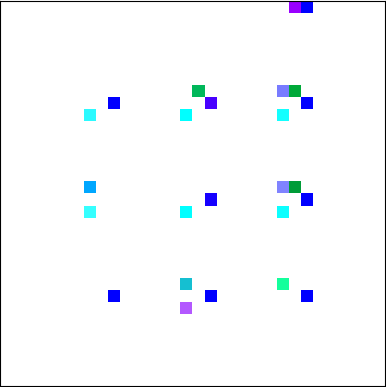}}\hfill
    \subfloat[Round 30]{\includegraphics[width=.18\textwidth]{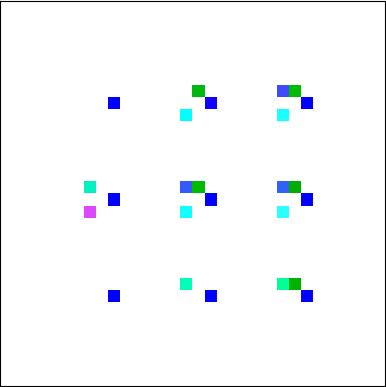}}\hfill
    \subfloat[Round 40]{\includegraphics[width=.18\textwidth]{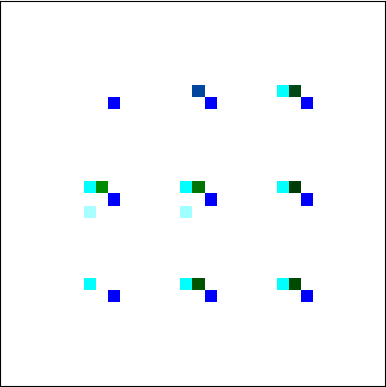}}\hfill
    \subfloat[Round 50]{\includegraphics[width=.18\textwidth]{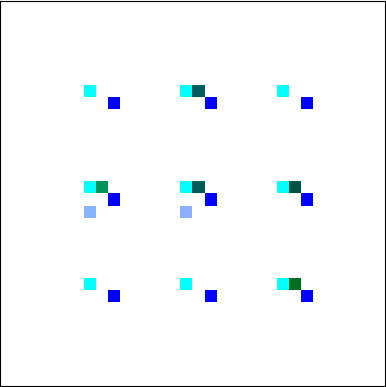}}\hfill
    \subfloat[Round 60]{\includegraphics[width=.18\textwidth]{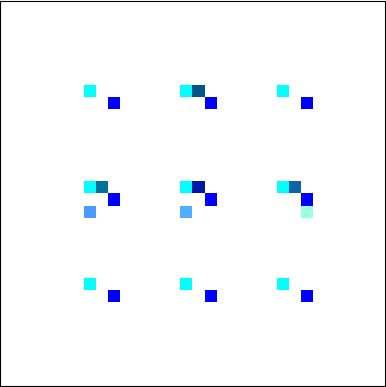}}\hfill
    \subfloat[Round 70]{\includegraphics[width=.18\textwidth]{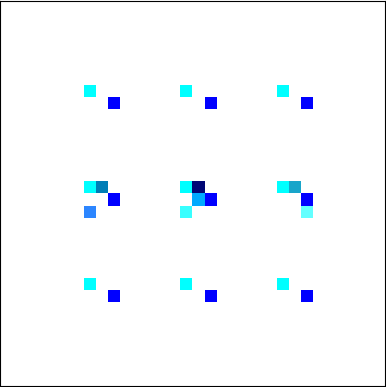}}\hfill
    \subfloat[Round 80]{\includegraphics[width=.18\textwidth]{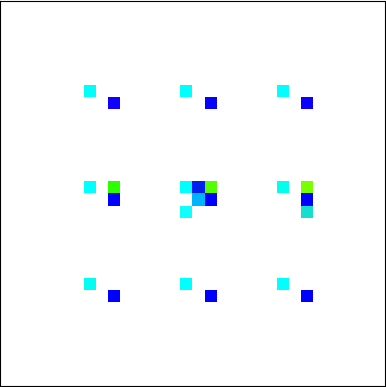}}\hfill
    \subfloat[Round 90]{\includegraphics[width=.18\textwidth]{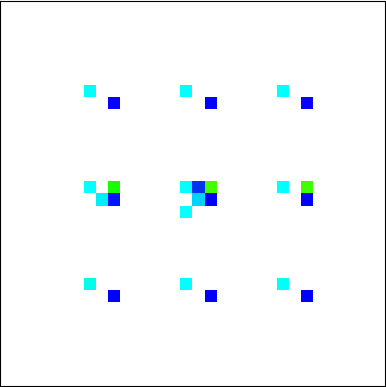}}\hfill
    \subfloat[Round 100]{\includegraphics[width=.18\textwidth]{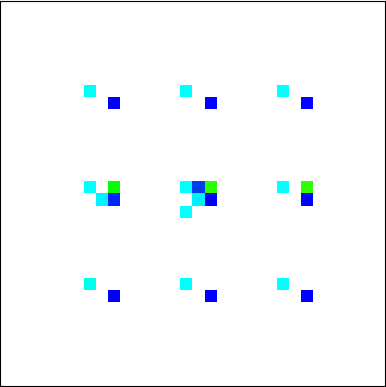}}\hfill
    \subfloat[Round 110]{\includegraphics[width=.18\textwidth]{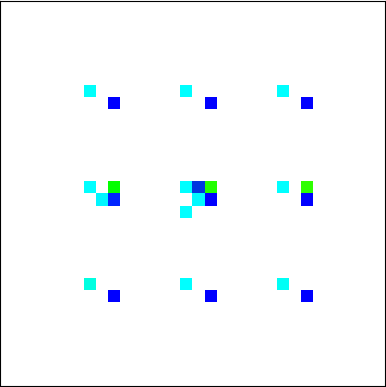}}\hfill
    \subfloat[Round 120]{\includegraphics[width=.18\textwidth]{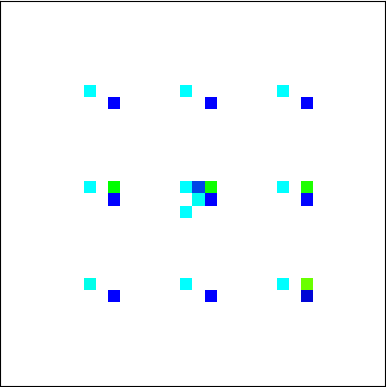}}\hfill
    \subfloat[Round 130]{\includegraphics[width=.18\textwidth]{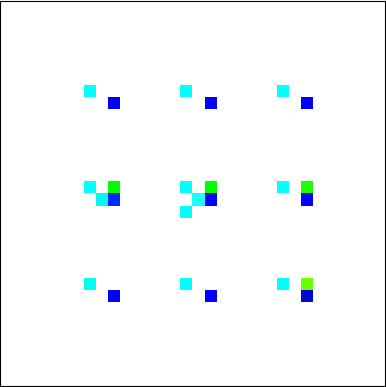}}\hfill
    \subfloat[Round 140]{\includegraphics[width=.18\textwidth]{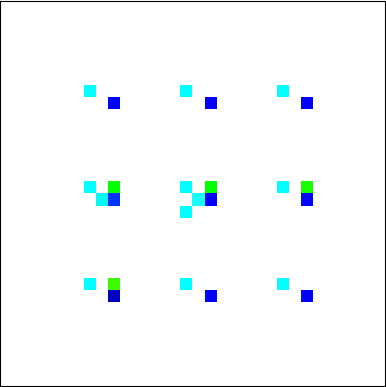}}\hfill
    \subfloat[Round 150]{\includegraphics[width=.18\textwidth]{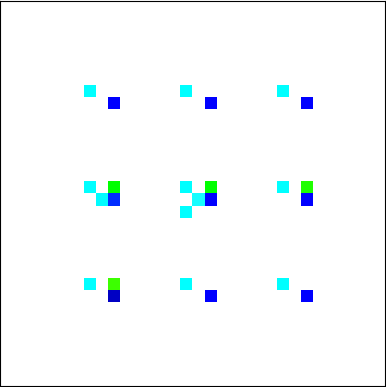}}
    \caption{(CIFAR10, ResNet18) \ourmodel{} triggers on different rounds.}
    \label{fig:cifar10_triggers}
\end{figure*}

\begin{figure*}[htbp]
    \centering
    \subfloat[Round 10]{\includegraphics[width=.18\textwidth]{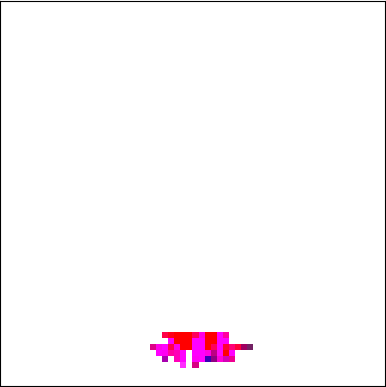}}\hfill
    \subfloat[Round 20]{\includegraphics[width=.18\textwidth]{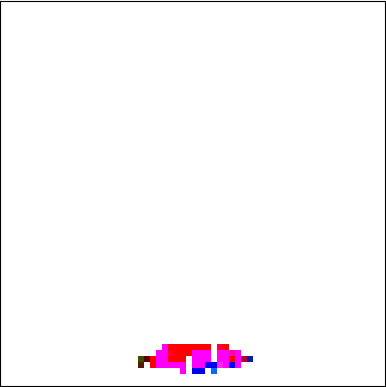}}\hfill
    \subfloat[Round 30]{\includegraphics[width=.18\textwidth]{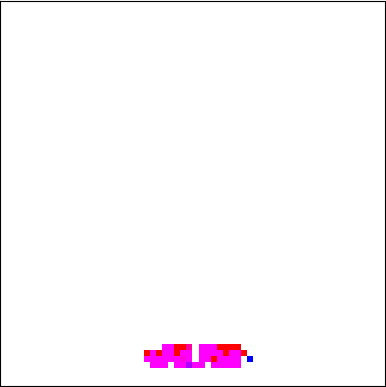}}\hfill
    \subfloat[Round 40]{\includegraphics[width=.18\textwidth]{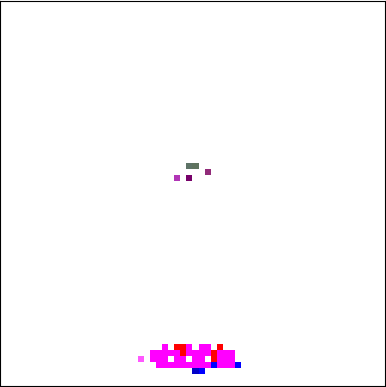}}\hfill
    \subfloat[Round 50]{\includegraphics[width=.18\textwidth]{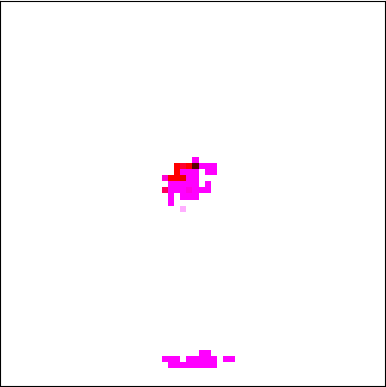}}\hfill
    \subfloat[Round 60]{\includegraphics[width=.18\textwidth]{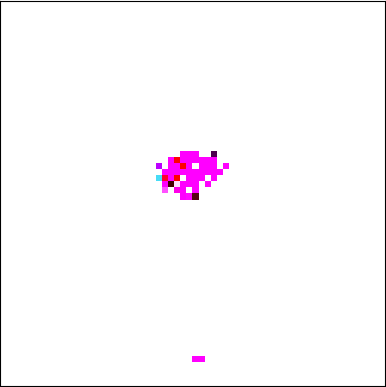}}\hfill
    \subfloat[Round 70]{\includegraphics[width=.18\textwidth]{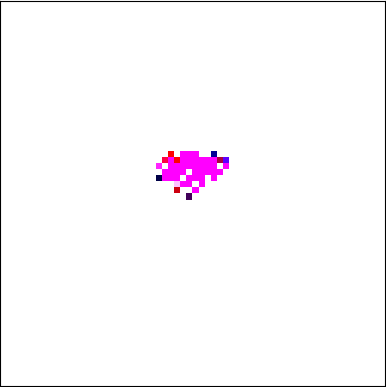}}\hfill
    \subfloat[Round 80]{\includegraphics[width=.18\textwidth]{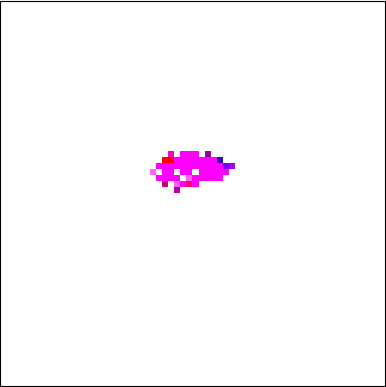}}\hfill
    \subfloat[Round 90]{\includegraphics[width=.18\textwidth]{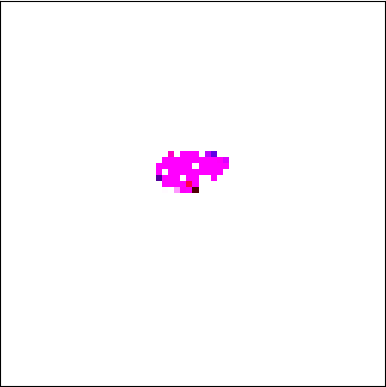}}\hfill
    \subfloat[Round 100]{\includegraphics[width=.18\textwidth]{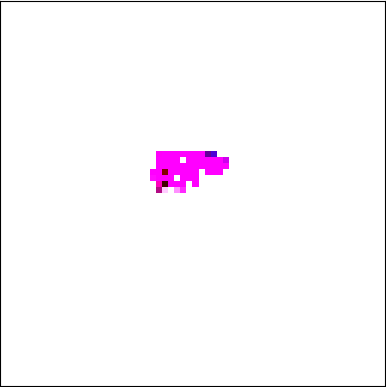}}\hfill
    \caption{(Tiny ImageNet, VGG11) \ourmodel{} triggers on different rounds.}
    \label{fig:tiny_triggers}
\end{figure*}

In Figure~\ref{fig:cifar10_triggers} and Figure~\ref{fig:tiny_triggers}, we demonstrated how \ourmodel{} trigger changes during the FL training. 

In Figure~\ref{fig:cifar10_triggers}, we showed one screenshot of the trigger on a blank background in the same size of the cifar10's figure for every ten global rounds.  These trigger screenshots were collected during a \ourmodel{} attacking experiment that trains ResNet18 as the global model on the CIFAR-10 dataset, with Trimmed Mean used as the aggregation rule. Figure~\ref{fig:cifar10_trimmean_plot} displays the Main-task Accuracy and Attack Success Rate of the global model over 150 global rounds in this experiment.

Similarly, in Figure~\ref{fig:tiny_triggers} we showed one screenshot of the trigger on a blank background in the same size of the Tiny ImageNet's figure for every ten global rounds. These trigger screenshots were collected during a \ourmodel{} attacking experiment that trains VGG11 as the global model on the Tiny ImageNet dataset, with Trimmed Mean used as the aggregation rule. Figure~\ref{fig:tiny_trimmean_plot} displays the Main-task Accuracy and Attack Success Rate of the global model over 100 global rounds in this experiment.

According to Figure~\ref{fig:cifar10_triggers} and Figure~\ref{fig:tiny_triggers}, the \ourmodel{} trigger does not change drastically over rounds; instead, it develops gradually and coherently. Since the \ourmodel{} trigger is optimized based on the global model's parameters, and the global model is in turn influenced by malicious model updates backdoored by the \ourmodel{} trigger, the \ourmodel{} trigger and the global model form a Markov chain. During training, as the global model evolves coherently and gradually, the states of the \ourmodel{} trigger evolve as well in the same pattern.

\begin{figure}[htbp]
    \begin{minipage}[c]{0.48\textwidth}
    \centering
    \begin{tikzpicture}
    \begin{axis}[
        title={Global model's accuracy on main/backdoor tasks},
        xlabel={Rounds},
        ylabel={Accuracy (\%)},
        xmin=-5, xmax=160,
        ymin=-5, ymax=105,
        xtick={0,30,60,90,120,150},
        ytick={0,20,40,60,80,100},
        legend pos=south east,
        grid=both,
        major grid style={line width=.2pt,draw=gray!50},
        minor grid style={line width=.1pt,draw=gray!20},
        width=\textwidth,
        height=0.6\textwidth,
        cycle list name=color list,
    ]
    
    \pgfplotstableread{./plot_data/cifar10_trimmean.dat}\datatable
    
    \addplot[color=blue, mark=none] table[x index=0, y index=1] {\datatable};
    \addlegendentry{MA}
    
    \addplot[color=red, mark=none] table[x index=0, y index=2] {\datatable};
    \addlegendentry{ASR}
    \end{axis}
    \end{tikzpicture}
\caption{Global model's accuracy in experiment of getting trigger screenshots in Figure~\ref{fig:cifar10_triggers}. (CIFAR10, ResNet18)}
\label{fig:cifar10_trimmean_plot}
\end{minipage}

\begin{minipage}[c]{0.48\textwidth}
    \centering
    \begin{tikzpicture}
    \begin{axis}[
        title={Global model's accuracy on main/backdoor tasks},
        xlabel={Rounds},
        ylabel={Accuracy (\%)},
        xmin=-5, xmax=110,
        ymin=-5, ymax=105,
        xtick={0,20,40,60,80,100},
        ytick={0,20,40,60,80,100},
        legend pos=south east,
        grid=both,
        major grid style={line width=.2pt,draw=gray!50},
        minor grid style={line width=.1pt,draw=gray!20},
        width=\textwidth,
        height=0.6\textwidth,
        cycle list name=color list,
    ]
    
    \pgfplotstableread{./plot_data/tiny_trimmean.dat}\datatable
    
    \addplot[color=blue, mark=none] table[x index=0, y index=1] {\datatable};
    \addlegendentry{MA}
    
    \addplot[color=red, mark=none] table[x index=0, y index=2] {\datatable};
    \addlegendentry{ASR}
    \end{axis}
    \end{tikzpicture}
\caption{Global model's accuracy in experiment of getting trigger screenshots in Figure~\ref{fig:tiny_triggers}. (Tiny ImageNet, VGG11)}
\label{fig:tiny_trimmean_plot}
\end{minipage}
\end{figure}

\clearpage

\section{Evaluation of \ourmodel{} attack against Flip~\cite{zhang2022flip}}\label{flip}

Flip~\cite{zhang2022flip} is a client-side defense strategy where benign clients perform trigger inversion and adversarial training using their local data to recover the global model from backdoors. In this section, we evaluate the effectiveness of the \ourmodel{} attack against the Flip defense. We implemented the \ourmodel{} attack by modifying the data preparation approach in Flip's open-source project, replacing it with the method used in this work, and injecting our data-poisoning algorithms into a subset of clients. Additionally, as \ourmodel{} is a pure data-poisoning attack, we removed any additional steps in their project specified to malicious clients but not existed in benign clients' training, to ensure consistency between malicious clients and benign clients in FL training. We selected Fashion MNIST as the main-task dataset for our evaluation and directly adopted Flip's default experiment settings provided in their project - the total number of clients was 100 and 4\% of them were malicious clients; the aggregation rule was set to FedAvg; the global model's parameters were initialized by a pre-trained state. The size of \ourmodel{} trigger was set to 64, consistent with our default attacking settings.

We compared the performance of the \ourmodel{} attack under two attack patterns provided by Flip's project: 1) \textbf{Single shot}: Each of the 4 malicious clients conducts a one-time attack at the beginning of training. 2) \textbf{Continuous}: All 4 malicious clients continuously execute the attack algorithms in every round during training.

Figure~\ref{fig:flip_asr_plot} shows the performance of the \ourmodel{} attack on an FL system using Flip as its defense, measured by the Attack Success Rate (ASR). In the single-shot attack pattern, \ourmodel{} maintains a stable ASR of around 15\% across all training rounds, exceeding the random guess accuracy of 10\% for the 10-class dataset. In the continuous attack pattern, \ourmodel{} achieves a significant ASR, peaking at 80.03\% during training and stabilizing around 40\%, which is higher than the single-shot pattern. These results indicate that Flip is vulnerable to optimized triggers with varying appearances across different rounds,  because recovering from backdoors is an after-effect strategy which is unable to stop new and distinct backdoors from injecting into the model.

Figure~\ref{fig:flip_ma_plot} illustrates the global model's performance on the main task data when using Flip as a defense while under \ourmodel{} attack. We observed that employing Flip reduces the global model's main-task performance compared to not using it. In our baseline experiment on Fashion MNIST, with the same data distribution and aggregation rule (FedAvg), the model achieved an 86.7\% MA. However, Flip's global model achieved only 82.8\% MA at its best by the end, even with pre-trained model initialization. Additionally, under continuous attack by the \ourmodel{} trigger, the global model's MA further declined compared to the less frequent attack pattern. This raises concerns about Flip's ability to maintain stable and normal performance on the main task while effectively defending against attacks.

\begin{figure}[htbp]
    \begin{minipage}[c]{0.48\textwidth}
    \centering
    \begin{tikzpicture}
    \begin{axis}[
        title={Global model's accuracy on backdoor task (\ourmodel{})},
        xlabel={Rounds},
        ylabel={Accuracy (\%)},
        xmin=-5, xmax=110,
        ymin=-5, ymax=105,
        xtick={0,20,40,60,80,100},
        ytick={0,20,40,60,80,100},
        legend pos=north east,
        grid=both,
        major grid style={line width=.2pt,draw=gray!50},
        minor grid style={line width=.1pt,draw=gray!20},
        width=\textwidth,
        height=0.6\textwidth,
        cycle list name=color list,
    ]
    
    \pgfplotstableread{./plot_data/flip_asr.dat}\datatable
    
    \addplot[color=blue, mark=none] table[x index=0, y index=1] {\datatable};
    \addlegendentry{Single shot}
    
    \addplot[color=red, mark=none] table[x index=0, y index=2] {\datatable};
    \addlegendentry{Continuous}
    \end{axis}
    \end{tikzpicture}
\caption{Global model's Attack Success Rate under \ourmodel{} attack when employed Flip as defense strategy. (Fashion MNIST) }
\label{fig:flip_asr_plot}
\end{minipage}

\begin{minipage}[c]{0.48\textwidth}
    \centering
    \begin{tikzpicture}
    \begin{axis}[
        title={Global model's accuracy on main task},
        xlabel={Rounds},
        ylabel={Accuracy (\%)},
        xmin=-5, xmax=110,
        ymin=65, ymax=85,
        xtick={0,20,40,60,80,100},
        ytick={65,70,75,80,85},
        legend pos=south east,
        grid=both,
        major grid style={line width=.2pt,draw=gray!50},
        minor grid style={line width=.1pt,draw=gray!20},
        width=\textwidth,
        height=0.6\textwidth,
        cycle list name=color list,
    ]
    
    \pgfplotstableread{./plot_data/flip_ma.dat}\datatable
    
    \addplot[color=blue, mark=none] table[x index=0, y index=1] {\datatable};
    \addlegendentry{Single shot}
    
    \addplot[color=red, mark=none] table[x index=0, y index=2] {\datatable};
    \addlegendentry{Continuous}
    \end{axis}
    \end{tikzpicture}
\caption{Global model's Main-task Accuracy under \ourmodel{} attack when employed Flip as defense strategy. (Fashion MNIST) }
\label{fig:flip_ma_plot}
\end{minipage}
\end{figure}

\section{Evaluation of \ourmodel{} attack against FRL~\cite{everyvotecounts}}\label{FRL}

FRL~\cite{everyvotecounts} is a defense strategy where the server sparsifies the value space of model updates, allowing clients to vote on the most effective model updates based on their local data. The server then aggregates only the accepted votes while rejecting outliers to construct the global model. In this section, we evaluate the effectiveness of the \ourmodel{} attack against the FRL defense. Similar to the experiment on Flip, we implemented our attack on FRL's open-source project by injecting our data-poisoning algorithms into a portion of clients' execution and removing any inconsistent steps that distinguished malicious clients from benign ones during training. We used FRL's default settings, in which only 2\% of clients were malicious, and tested our attack on the CIFAR10 dataset as the main training task.

Table~\ref{tbl:FRL} presents the performance results of the \ourmodel{} attack on an FL system employing FRL as the defense method. The ASR of \ourmodel{} (92.5\%) is significantly higher than that of other backdoor attack approaches tested and discussed in FRL's paper. This indicates that FRL, which relies on analyzing clients' model updates, is vulnerable to our attack. The evaluation results also demonstrate that the \ourmodel{} attack is more advanced than backdoor attacks with static objectives when targeting the FRL defense strategy.

\begin{table}[hbpt]
    \renewcommand{\arraystretch}{1.1}
    \addtolength{\tabcolsep}{-1pt}
    \begin{center}
      \begin{tabular}{|c|c|}
        \hline
        Attacks & ASR \\
        \hline
        Semantic backdoor attacks & 49.2 \\
        \hline
        Artificial backdoor attacks & 0 \\
        \hline
        Edge-Case backdoor attacks & 64.6 \\
        \hline
        \textbf{\ourmodel{} backdoor attacks} & \textbf{92.5}\\
        \hline
      \end{tabular}
    \end{center}
    \vspace{-4mm}
    \caption{Comparison results on CIFAR10.}
    \label{tbl:FRL}
    \vspace{-3mm}
  \end{table}

\vspace{-2mm}

\section{Effects of the scaling-based model poisoning techniques on attacks}
\begin{figure}[htbp]
  \vspace{-4mm}
  \begin{subfigure}[htbp]{0.48\linewidth}
    \centering
    \caption{Final $\mathit{ASR}$}
    \label{fig:scale_final}
  \begin{tikzpicture}
  \begin{axis}[
  xmode=log,
  log basis x={2}, 
  ytick={0,20,40,60,80,100},
  ymin=0,
  ymax=105,
  xtick={0.5, 1, 2, 8, 32, 128}, 
  ymajorgrids,
  height=4.5cm,width=5cm,                                
  xlabel = Scaling Factor,
  xticklabel style={text height=1.5ex}, 
  legend style={at={(1,1)},legend columns=1, font=\footnotesize},
  ] 
  \addplot[
    color=black,
    mark=triangle,
    mark options = {solid},
    semithick,
    ]
    coordinates {
    (0.5,57.65)(1,99.23)(3,23.13)(9,39.50)(33,37.34)(129,26.53)
    }; 
  
  \addplot[
    color=red,
    mark=x,
    mark options = {solid,semithick},
    densely dashed
    ]
    coordinates {
    (0.5,38.17)(1,92.60)(3,1.94)(9,3.21)(33,1.22)(129,5.78)
    }; 
  \addplot[
    color=blue,
    mark=+,
    mark options = {solid,semithick},
    densely dashdotted
    ]
    coordinates {
    (0.5,3.70)(1,99.95)(3,4.79)(9,3.23)(33,2.05)(129,2.79)
    }; 
    \legend{Ours, FT, DFT}
  
  \end{axis} 
  \end{tikzpicture}
  \end{subfigure}
\hfill
  \begin{subfigure}[htbp]{0.5\linewidth}
    \centering
    \caption{Average $\mathit{ASR}$}
    \label{fig:scale_avg}
    \begin{tikzpicture}
    \begin{axis}[
    xmode=log,
    log basis x={2}, 
    ytick={0,20,40,60,80,100},
    ymin=0,
    ymax=105,
    xtick={0.5, 1, 2, 8, 32, 128}, 
    ymajorgrids,
    height=4.5cm,width=5cm,                                
    xlabel = Scaling Factor,
    xticklabel style={text height=1.5ex}, 
    legend style={at={(1,1)},legend columns=1, font=\footnotesize},
    ] 
    \addplot[
      color=black,
      mark=triangle,
      mark options={solid},
      semithick,
      ]
      coordinates {
      (0.5,29.05)(1,86.15)(3,17.40)(9,32.08)(33,31.92)(129,22.63)
      }; 
    
    \addplot[
      color=red,
      mark=x,
      mark options={solid,semithick},
      densely dashed
      ]
      coordinates {
      (0.5,31.39)(1,82.94)(3,3.31)(9,4.37)(33,0.86)(129,4.23)
      }; 
    \addplot[
      color=blue,
      mark=+,
      mark options={solid,semithick},
      densely dashdotted
      ]
      coordinates {
      (0.5,3.32)(1,78.96)(3,6.26)(9,2.98)(33,3.37)(129,3.48)
      }; 
      \legend{Ours, FT, DFT}
    
    \end{axis} 
    \end{tikzpicture}
  \end{subfigure}  
  \vspace{-2mm}
  \caption{ Comparison results of different attacks when employing the scaling-based model poisoning technique to undermine FLAME defense (implemented on the FEMNIST dataset).}
  \label{fig:scale}
  \vspace{-4mm}
  \end{figure}

  In this section, we removed the TEEs assumption and conducted experiments to examine the effects of employing scaling-based model poisoning techniques on the attack performance of \ourmodel{}, FT, and DFT. By incorporating the model poisoning technique, our implementation of FT and DFT attacks aligns more closely with the attack strategies introduced in state-of-the-art backdoor attacks on FL \cite{howtobackdoor,Xie2020DBA:}.

  Our experiments were designed within an FL system utilizing FLAME as its aggregation rule and FEMNIST dataset as its main training task. We adjusted the scaling factors, used to scale malicious clients' model updates, to be 0.5, 1, 3, 9, 33, and 129 respectively. Figures \ref{fig:scale_final} and \ref{fig:scale_avg} illustrate the results of Final $ASR$ and Avg $ASR$ of various attacks in response to different scaling factors.
  
  We observed that when the scaling factor is 1, all \ourmodel{}, FT, and DFT attacks exhibit comparable and high $ASR$ against FLAME defense. However, as the scaling factor increases, FLAME demonstrates robust defense performance, significantly reducing the $ASR$ of every attack pipeline. Despite this mitigation, \ourmodel{} shows greater resilience in attack effectiveness compared to FT and DFT. The optimized trigger generated by our algorithms retains intrinsic attack effects on the global model even without successful data-poisoning techniques. When the scaling factor is reduced to 0.5, malicious model updates are expected to be stealthier, yet their contributions to the aggregated global model are also mitigated, resulting in reduced $ASR$ for all attacks compared to when the scaling factor is 1.

\section{Main-task Accuracy Results}

Table~\ref{tbl:main_MA} lists the Main-task Accuracy of each experiment in getting results in Figure~\ref{fig:main_ASR}. Table~\ref{tbl:main_MA} demonstrates that for different datasets used as the main tasks, global models under various attacks maintained a comparable level of Main-task Accuracy to the baselines with no attacks (``None''), indicating that all types of backdoor attacks successfully achieved their stealthiness goals.

Table \ref{tbl:MCR_MA_CIFAR10} lists the global model's Main-task Accuracy of each experiment in getting results in Table~\ref{tbl:MCR_ASR_CIFAR10}.  Table~\ref{tbl:MCR_ASR_CIFAR10} evaluates the impact of different malicious client ratios on the attack effectiveness of various attacks when using the CIFAR10 as the main-task dataset.  Table~\ref{tbl:MCR_MA_CIFAR10} demonstrates that the performance of global models on Main-task data is not affected by changes in the malicious client ratio, indicating that the stealthiness goals of all backdoor attacks were achieved. “None” represents the baseline MA results with no attack present during FL training.

Table~\ref{tbl:TS_MA_CIFAR10} lists the global model's Main-task Accuracy of each experiment in getting results in Table~\ref{tbl:TS_ASR_CIFAR10}. Table~\ref{tbl:TS_ASR_CIFAR10} evaluates the impact of different trigger sizes on the attack effectiveness of various attacks when using the CIFAR10 as the main-task dataset. Table~\ref{tbl:TS_MA_CIFAR10} demonstrates that the performance of global models on Main-task data is not affected by changes in the trigger sizes, indicating that the stealthiness goals of all backdoor attacks were achieved. ``None'' represents the baseline MA results with no attack present during FL training.

\begin{table*}[h]
    \vspace{-2mm}
    \renewcommand{\arraystretch}{1}
    \centering
    \small\addtolength{\tabcolsep}{-2pt}
    \begin{tabular}{|c|cccc|cccc|cccc|cccc|}
    \hline
    \multirow{2}{*}{\textbf{MA}} &\multicolumn{4}{c|}{\textbf{Tiny ImageNet}} & \multicolumn{4}{c|}{\textbf{Fashion MNIST}} & \multicolumn{4}{c|}{\textbf{FEMNIST}}  &\multicolumn{4}{c|}{\textbf{CIFAR10}}\\
    \cline{2-17} 
    & None & Ours & FT & DFT& None & Ours & FT & DFT& None & Ours & FT & DFT& None & Ours & FT & DFT\\
    \hline
    FedAvg & 43.9 & 43.5	& 43.0 & 43.3 &86.7 &87.3&86.7&86.8& 82.2&81.4&83.3&82.3& 70.3&70.7&70.4&71.4\\
    \hline
    Median & 40.6 &40.2 &40.6 &38.6& 86.0&85.8&86.6&86.3& 80.4&81.5&79.8	&79.9& 70.2&69.1&69.8&69.7\\
    \hline
    {\footnotesize Trimmed Mean} & 40.8 &40.4	&40.1	&40.6& 86.4&85.8&86.4&86.3& 80.2&81.7&81.3&81.2& 69.4&70.4&70.2&70.8\\
    \hline
    RobustLR & 44.1 &42.7&42.9&43.2 & 86.5 &86.8&86.6&86.9&81.8 & 82.5&81.9&82.6& 70.4 &70.1&70.3&70.5\\
    \hline
    RFA & 43.6 &43.0&43.0&43.0&86.4 &86.0&87.1&87.1& 83.0& 80.7& 81.0& 80.8&70.4 &70.7&70.3&70.8\\
    \hline
    FLAIR & 43.6 &42.6&41.8&42.1  &86.1 &84.9&85.2&84.4  & 81.5&80.7&80.6&79.7  &70.3 &70.6&71.0&70.4\\
    \hline
    FLCert &40.3 &40.2&39.7&39.7& 86.2&85.9&86.0&86.8& 81.3&80.9&	81.5&	81.0& 69.6&70.0&69.8&70.4\\
    \hline
    FLAME &29.9 &28.7&29.2&28.9& 86.4&86.4&86.4&86.7& 81.8&80.2&80.7&81.0& 70.1&70.3	&70.9&70.9\\
    \hline
    FoolsGold &43.1 &43.2&43.5&43.2& 86.6&87.1	&86.8&87.3& 83.4&82.7&83.0&81.8& 70.4&71.0&71.2&71.7\\
    \hline
    Multi-Krum &30.7 &27.7&27.7&26.4& 86.2&85.9&86.0&87.0&79.9 &80.4&79.6&80.2& 61.4&63.0&63.2&60.8\\
    \hline
  
    \end{tabular}
    \caption{The Main-task Accuracy (MA) of global models in getting representative results in Figure~\ref{fig:main_ASR}. "None" represents no attack existing in the FL training.}
    \label{tbl:main_MA}
    \vspace{-5mm}
    \end{table*}

\begin{table*}[h]
    \centering
    \small
    \begin{tabular}{|c|c|ccc|ccc|ccc|ccc|}
        \hline
         \textbf{MCR} & &\multicolumn{3}{c|}{\textbf{0.05}} & \multicolumn{3}{c|}{\textbf{0.1}} & \multicolumn{3}{c|}{\textbf{0.2}}  &\multicolumn{3}{c|}{\textbf{0.3}}\\
        \hline
        & None & Ours & FT & DFT&  Ours & FT & DFT&  Ours & FT & DFT&  Ours & FT & DFT\\
        \hline
    FedAvg  &70.3 &70.66 &70.37 &71.37 &70.03 &71.04 &70.13 &69.9 &70.39 &71.18 &70.25 &70.69 &70.24\\
    \hline
    Median  &70.21 &69.06 &69.76 &69.71 &69.32 &69.17 &70.12 &68.23 &69.05 &68.87 &68.49 &68.47 &67.82\\
    \hline
    {\footnotesize Trimmed Mean}  &69.43 &70.42 &70.24 &70.84 &69.9 &69.17 &69.78 &69.33 &69.19 &69.8 &69.23 &68.83 &68.02\\
    \hline
    RobustLR & 70.35 &70.10 &70.35 &70.48 &70.58 &70.42 &69.90 &70.31 &70.56 &70.43 &70.05 &69.11 &69.22  \\
    \hline
    RFA  &70.42  &70.69 &70.27 &70.77 &70.35 &70.44 &70.16 &70.72 &70.33 &69.56 &70.09 &69.72 &69.37    \\
    \hline
    FLAIR  & 70.25 &70.62 &71.04 &70.42 &69.80 &71.45 &70.89 &71.85 &71.20 &71.16 &71.26 &69.74 &70.99\\
    \hline
    FLCert  &69.6 &69.95 &69.76 &70.42 &69.44 &69.44 &69.45 &69.28 &69.25 &69.73 &68.54 &69.06 &68.24\\
    \hline
    FLAME  &70.14 &70.28 &70.93 &70.85 &69.62 &70.87 &71.01 &70.71 &70.4 &70.58 &69.19 &71.45 &70.52\\
    \hline
    FoolsGold  &70.42 &71.02 &71.19 &71.68 &70.71 &71.32 &71.27 &70.45 &70.38 &70.82 &70.12 &69.97 &69.97\\
    \hline
    Multi-Krum  & 61.38 &62.98 &63.16 &60.80 &61.44 &62.89 &62.09 &59.38 &61.26 &63.70 &60.28 &64.02 &62.96\\
    \hline 
    \end{tabular}
    \caption{The Main-task Accuracy (MA) of global models under different attacks at varying malicious client ratios. (CIFAR10).}
    \label{tbl:MCR_MA_CIFAR10}
    \end{table*}

\begin{table*}[h]
    \centering
    \small
    \begin{tabular}{|c|c|ccc|ccc|ccc|ccc|}
        \hline
          \textbf{Trigger Size}& &\multicolumn{3}{c|}{\textbf{9}} & \multicolumn{3}{c|}{\textbf{25}} & \multicolumn{3}{c|}{\textbf{49}}  &\multicolumn{3}{c|}{\textbf{100}}\\
        \hline
        & None & Ours & FT & DFT&  Ours & FT & DFT&  Ours & FT & DFT&  Ours & FT & DFT\\
        \hline
    FedAvg & 70.3 &70.88 &70.72 &71.25 &70.66 &70.37 &71.37 &70.77 &71.35 &70.94 &69.92 &70.71 &71.15\\
    \hline
    Median & 70.21 &68.31 &70.04 &68.69 &69.06 &69.76 &69.71 &69.95 &70.54 &70.56 &69.88 &70.30 &70.86\\
    \hline
    {\footnotesize Trimmed Mean} & 69.43 &69.75 &70.13 &70.19 &70.42 &70.24 &70.84 &69.42 &70.17 &69.79 &69.67 &70.26 &70.68\\
    \hline
    RobustLR  & 70.35 &70.48 &70.95 &69.48 &70.10 &70.35 &70.48 &70.79 &70.08 &70.27 &70.39 &69.73 &69.86 \\
    \hline
    RFA  & 70.42 &70.45 &70.16 &71.00 &70.69 &70.27 &70.77 &70.56 &70.19 &70.62 &70.52 &69.22 &70.77    \\
    \hline
    FLAIR &70.25  &70.79 &70.67 &70.58 &70.62 &71.04 &70.42 &70.84 &69.96 &71.03 &71.17 &70.65 &70.28\\
    \hline
    FLCert & 69.6 &69.88 &69.64 &69.87 &69.95 &69.76 &70.42 &67.77 &69.83 &70.08 &68.81 &70.81 &70.41\\
    \hline
    FLAME & 70.14 &70.07 &71.24 &70.19 &70.28 &70.93 &70.85 &69.87 &71.20 &70.68 &67.24 &71.06 &70.75\\
    \hline
    FoolsGold & 70.42 &70.4 &72.1 &70.09 &71.02 &71.19 &71.68 &70.66 &70.75 &71.38 &69.84 &71.06 &71.64\\
    \hline
    Multi-Krum & 61.38 &62.86 &64.65 &58.90 &62.98 &63.16 &60.80 &58.23 &60.16 &64.04 &63.03 &61.64 &63.33\\
    \hline
    \end{tabular}
    \caption{The Main-task Accuracy (MA) of global models under different attacks with varying trigger sizes. (CIFAR10).}
    \label{tbl:TS_MA_CIFAR10}
\end{table*}

\clearpage

\end{document}